\documentclass[journal,comscoc]{IEEEtran}
\usepackage{amsmath}
\usepackage{amssymb}
\usepackage{cite}
\usepackage{graphicx} 
\usepackage{multicol}
\usepackage{epstopdf}
\usepackage{verbatim}
\usepackage{url}
\usepackage{amsmath}
\usepackage{breqn}
\usepackage{amsthm}
\usepackage{multirow}
\usepackage{lipsum}
\usepackage{bbm}
\usepackage{caption}
\usepackage{algorithm}
\usepackage{algpseudocode}
\usepackage{mathtools}

\newtheorem{thm}{Theorem}
\newtheorem{lemma}{Lemma}

\newtheorem{defn}{Definition}

\newtheorem{remark}{Remark}
\usepackage{dblfloatfix}
\usepackage{array}
\ifCLASSOPTIONcompsoc
\usepackage[caption=false,font=footnotesize,labelfont=sf,textfont=sf]{subfig}
\else
\usepackage[caption=false,font=footnotesize]{subfig}
\fi

\usepackage{enumitem}

\begin{document}
\title{Distributed Learning Algorithms for Opportunistic Spectrum Access in Infrastructure-less Networks}
\author{Rohit Kumar, Sumit J. Darak, Manjesh K. Hanawal and ~Ankit~Yadav    
\thanks{Rohit Kumar is with ECE Dept., NIT Delhi, India. E-mail: rohitkumar@nitdelhi.ac.in.}
\thanks{Sumit J. Darak is with the ECE Dept., IIIT Delhi, Delhi, India}
\thanks{Manjesh K. Hanawal is with IEOR, IIT Bombay, India.}
\thanks{Ankit Yadav is with Electrical Engg. Dept., Texas, A \& M University, USA.}}

%
%
%

\maketitle
\begin{abstract}
An opportunistic spectrum access (OSA) for the infrastructure-less (or cognitive ad-hoc) network has received significant attention thanks to emerging paradigms such as the Internet of Things (IoTs) and smart grids. Research in this area has evolved from the $\rho^{rand}$ algorithm requiring prior knowledge of the number of active secondary users (SUs) to the musical chair (MC) algorithm where the number of SUs are unknown and estimated independently at each SU. These works ignore the number of collisions in the network leading to wastage of power and bring down the effective life of battery operated SUs. In this paper, we develop algorithms for OSA that learn faster and incurs fewer number of collisions i.e. energy efficient. We consider two types of infrastructure-less decentralized networks: 1) static network where the number of SUs are fixed but unknown, and 2) dynamic network where SUs can independently enter or leave the network. We set up the problem as a multi-player mult-armed bandit and develop two distributed algorithms. The analysis shows that when all the SUs independently implement the proposed algorithms, the loss in throughput compared to the optimal throughput, i.e. regret, is a constant with high probability and significantly outperforms existing algorithms both in terms of regret and number of collisions. Fewer collisions make them ideally suitable for battery operated SU terminals. We validate our claims through exhaustive simulated experiments as well as through a realistic USRP based experiments in a real radio environment.
\end{abstract}


\section{Introduction}
\IEEEPARstart{T}{he} need to increase the utilization of an electromagnetic spectrum has always been a concern for the service operators. With emerging paradigms such as the Internet of Things (IoTs) and smart grids consisting of thousands of devices transmitting intermittently, it will be difficult to follow static spectrum allocation policies due to limited spectrum below 6 GHz and high spectrum costs \cite{p1,ca1,ca2}. Hence, industry as well as academia are exploring various approaches such as opportunistic spectrum access (OSA), device-to-device communications, cellular-to-WiFi offloading and LTE-unlicensed (LTE-U) to meet the spectrum needs of these paradigms \cite{p1,p2,p3,ca1,ca2}. Among them, OSA based cognitive radio network (CRN) seems to be a promising solution which enables the devices to identify and exploit different part of the spectrum depending on the availability, type of service and device capabilities. DARPA's spectrum collaboration challenge 2016 was a significant step to bring CRN to life \cite{darpa}. Recently, 3GPPP new radio (NR) specifications for 5G confirm the use of DSA for operations in the shared and unlicensed spectrum. 
	
The CRN consists of licensed or primary users (PU) and unlicensed or secondary users (SU). In this paper, we consider OSA in the infrastructure-less overlay CRN (or cognitive radio ad-hoc networks \cite{ca1,ca2}) where PUs coordinate for orthogonal channel assignments through base stations while such coordination is not feasible for SUs. Also, SUs need to sense the channel for the presence of PUs since they can transmit only if the channel is idle. Transmissions of the SUs are time-slotted and packeted with acknowledgment from the receiver for received packet(s). We assume that for each channel idle process is independently and identically distributed across the time slots and independent of the other channels\footnote{More realistic 
	Markovian channel behavior can also be studied using Multi-Armed Bandit for Markov Chains. We leave it for future work.}. Due to the lack of coordination, multiple SUs may transmit on the same idle channel leading to a collision. The collisions not only lead to the loss in throughput but also result in wastage of battery due to retransmissions. The OSA becomes more challenging in dynamic networks where the SUs can enter or leave the network anytime without prior agreement. Such network are also referred to as cognitive ad-hoc and sensor networks consisting of hundreds of transmitting devices (analogous to SUs), but only a few of them are active at a time \cite{ca1,ca2}. We develop algorithms to enable collision-free communication between such devices without the need of any control channel or coordination via the central controller.

To overcome the lack of coordination among SUs, several distributed algorithms \cite{MC,MEGA,quek,prand,rjain,sdsp,seus,zhandi,zhandia,
	tdfs,gai1,gai2,zandi,kaufmann} are proposed which guarantee orthogonal channel allocations if faithfully implemented by all the SUs. Existing algorithms assume prior knowledge of the number of active SUs ($U$) in the network. To the best of our knowledge, algorithms in \cite{MEGA} and \cite{MC} are the only ones that are agnostic to the number of active SUs ($U$) in the network. However, these algorithms estimate $U$ based on the number of collisions observed in the network. They force the SUs to randomly select the channels in the initial phase so that a large number of collisions are observed by each SU. This results in significant loss of throughput and also wastage of transmission power as each collision results in reprocessing and retransmission of the same (lost) packet. Hence these algorithms are not suitable for battery operated devices which are power constrained. Our goal in this work is to develop distributed algorithms for OSA that offer better throughout with a negligible number of collisions than that offered by the state-of-the-art algorithms. 

	The total throughput for the SUs is highest if all of them select orthogonal channels from the top $U$ channels\footnote{the '$U$ top channels' refers to the set of first $U$ channels when arranged in the decreasing order of their probability of being idle. The top channel is the channel with the highest probability of being idle}. The existing distributed algorithms thus aim to learn the channel statistics as well as the number of SUs (if unknown) and then find orthogonal channel assignments in the top $U$ channels. In this paper, we demonstrate that knowing $U$ is not necessary to find orthogonal channel allocation in the top channels once all the SUs learn the channel statistics. Specifically, we develop algorithms based on novel trekking approach where SUs operating on a channel always looks to operate on a channel with the better probability of being idle if no other SU is operating on that channel. Thus, all the SUs end up transmitting on the top channels without knowing how many SUs are there in the network. 
	
	In dynamic networks, existing algorithms follow epoch approach where SUs reestimate the channel characteristics and $U$ at the beginning of each epoch leading to further degradation in performance. The proposed continuous trekking based approach for dynamic networks guarantee maximum utilization of top channels without the need of deterministic or epoch approach, and any restrictions on SUs movement. In case of both algorithms, we guarantee 'fairness' in channel allocations over multiple experiments. 
	
	The proposed algorithms minimize the regret in a multi-player multi-armed bandits where regret is defined as the difference between the best aggregate throughput achievable when all the SUs cooperate with prior knowledge of network parameters (channel statistic and number of SUs) and the throughput achieved without coordination and any prior knowledge of the network parameters. Our contributions can be summarized as follows:

	\begin{enumerate}
	
\item 
For OSA in static networks with fixed but unknown number of SUs, we propose algorithm TSN (Trekking for Static Networks) and show that it gives constant regret with high confidence.

\item
For OSA in dynamic networks where SUs can enter or leave the network any time, we propose algorithm TDN (Trekking for Dynamic Networks) and show that it gives $O(\sqrt{T})$ regret with high confidence.

\item 
We validate our algorithms through extensive simulations which show their superiority over existing algorithms. 
\item We give a realistic universal software radio peripheral (USRP) based experimental setup and demonstrate the effectiveness of our algorithms in a real radio environment. 

\end{enumerate}

This paper is a significant extension of \cite{wiopt} in which we present TSN algorithm. Here, we provide theoretical bounds for the regret and number of collisions for the TSN algorithm, and validation via simulation as well as experimental results. We also present TDN algorithm for dynamic networks and its analysis. The rest of the paper is organized as follows: In Section~\ref{prevwork}, we present the literature review followed by the network model in Section~\ref{network_model}. Section~\ref{SN} and Section~\ref{DN} describe the proposed TSN and TDN algorithms, respectively along with their performance analysis. Section \ref{sim_res} offers discussion on the synthetic results followed by experimental results in Section \ref{exp_res}. Section \ref{conclusion} concludes the paper.
	
\vspace{-0.2cm}
\section{Literature Review}\label{prevwork}
In this section, we review some of the recent works related to OSA. For OSA in cooperative networks, various algorithms have been proposed using deterministic or auction based approaches for SU orthogonalization in top channels \cite{rjain,zhandia}. However, for collision-free transmissions, they either need a central controller or communication links between SUs. The lack of both makes the OSA in an infrastructure-less CRN a difficult and challenging problem. Here, we limit the discussion to the papers related to this domain.
	
	The  time division fair sharing (TDFS) \cite{tdfs} and $\rho^{rand}$ \cite{prand}, are the first works which enable SU orthogonalization in top $U$ channels in the infrastructure-less network. Both algorithms employ upper confidence bound (UCB) based multi-armed bandit (MAB) algorithm for characterization of channels and randomization-based rank selection for SU orthogonalization. Though both algorithms offer identical regret, $\rho^{rand}$ \cite{prand} is a preferred choice when frequency band switching cost is high. The algorithms in \cite{gai1,gai2,sdsp,kaufmann} extend $\rho^{rand}$ using other MAB algorithms such as UCB extensions, Thompson Sampling and Bayesian UCB to improve the regret, and frequency band switchings. To reduce the number of collisions in $\rho^{rand}$, \cite{zandi} uses a larger subset of channels than top $U$ channels during exploration phase while algorithms in \cite{seus,zhandi} replace randomization approach for rank selection in $\rho^{rand}$ with the MAB based learning approach.  Though all these algorithms \cite{prand,tdfs,sdsp,gai1,gai2,zhandi,seus,zandi,kaufmann} offer lower regret and fewer number of SU collisions than random channel selection approach, they may not be suitable for battery operated SUs in the infrastructure-less CRN since they need prior knowledge of number of active SUs, $U$ and use computationally intensive MAB algorithms. Another major drawback of these algorithms is that they assume the static network with a fixed number of SUs.
	
	    The algorithm in \cite{quek} is based on two-stage sequential channel hopping and does not need prior knowledge of $U$ for SU orthogonalization. Furthermore, it guarantees the negligible number of SU collisions. The drawback is that the SUs select all channels uniformly leading to high regret. The MEGA \cite{MEGA} and MC \cite{MC} are the only algorithms which do not need prior knowledge of $U$. It has been shown in \cite{MC} that the MC algorithm outperforms MEGA algorithm and is computationally efficient. The MC algorithm divides the time horizon into two stages: 1)~Learning stage, 2) MC stage. In the learning stage, each SU randomly chooses a channel in each time slot and observes the throughput as well as the number of collisions on them. This information is then exploited to estimate the number of active SUs in the network and orthogonalize SUs in one of the top channels in the MC stage. The MC algorithm in \cite{MC} is designed to work in an unlicensed spectrum where there is no PUs, and its extension for licensed spectrum has been discussed in \cite{ICL}. The MC algorithm incurs a significant number of collisions before it learns the number of SUs in the network. This leads to inefficient usage of battery power, spectrum and time. Another drawback is that it follows epoch approach for the dynamic networks. In this approach, SU resets MC algorithm at the beginning of each epoch and hence, needs learning stage in each epoch for re-estimation of $U$ leading to higher regret and collisions \cite{MC}. Also, epoch approach prohibits the entry or exit of SUs during each learning stage. In addition, to know the status of the horizon and epoch, inactive SUs either need to remain connected to the network instead of sleep mode or central controller is needed to convey the horizon status.

	 To the best of our knowledge, existing algorithms except MC and MEGA require prior knowledge of network parameters to achieve lower regret in the infrastructure-less decentralized CRN. This paper aims to develop algorithms that overcome these limitations while considering a futuristic and realistic network with no control channel for SUs. Further, we consider the dynamic network with no restriction on the movement (Entry or Exit) of the SUs. Hence, our performance guarantees are pessimistic. They can be improved if some time-bound restrictions are imposed.

\vspace{-0.2cm}
\section{Network Model}\label{network_model}
\vspace{-0.055cm}
In this section, we present the well-known network model for infrastructure-less (or ad-hoc) CRN which has been considered in many recent works including \cite{ca1,ca2,MEGA,quek,prand,rjain,sdsp,seus,zhandi,zhandia,
	tdfs,gai1,gai2,zandi,kaufmann}. 
It consists of $U$ SUs and $N$ channels in the  wideband licensed spectrum such that $N \geq U$. We assume time slotted communication where the horizon is divided into $T$ number of time slots, i.e., $t \in \{1,2,..,T\}$. The status of the channel in any slot $t$ can be either vacant or occupied. Each time slot is divided into two sub-slots. In the first sub-slot, each SU senses the channel for active PUs. For simplicity of analysis, we assume ideal detector, i.e., no sensing error. In the second sub-slot, they transmit if the channel is vacant. When more than one SU transmit on the same vacant channel, a collision occurs. If no collision occurs, data transmission is considered to be successful.

We assume that the channel $i$ being vacant is governed by some mean $\mu_i\in$ (0,1], $i\in[N]$ which is unknown to SUs and assumed to be independently and identically distributed across time slots. 
Let $\mu=[\mu_1, \mu_2,...,\mu_N]$ denotes the channel availability statistics. Without the loss of generality, we assume that $\mu_1 > \mu_2 >... >\mu_N$. Let $\mu_{min} = \min\limits_{i} \mu_i$. We assume $\mu_{min}>\theta>0$, otherwise some channels will never be vacant. This assumption also implies that $\sum\limits_{i = 1}^{N} \frac{\mu_i}{N} > \theta$. For later use, define $\Delta_i=\mu_{i+1}-\mu_{i},i=1,2,\ldots,N-1$ i.e., gap between $i^{th}$ and $(i+1)^{th}$ channel statistics and assume $\Delta_i>0$. Similar assumptions have been made in many recent works \cite{ca1,ca2,MEGA,quek,prand,zhandi,tdfs,gai1,gai2,zandi}.

%

	 We evaluate the performance of our algorithms in terms of expected regret defined as the difference between expected optimal throughput and runtime average throughput given as:  
	 \begin{align} \label{eq:regret} 
	 R_T &= R_{op}-E\left[\sum\limits_{t=1}^{T}\sum\limits_{u=1}^{U} r^u_t\right] \nonumber \\
	 &= R_{op}-\sum\limits_{t=1}^{T}\sum\limits_{u=1}^{U} \mu_{I_t^u}(1- E\left[C_{I_t^u}^u\right]).	 
	 \end{align}
where $R_{op}$ is the maximum mean total throughput achievable for SUs. It is achieved when each SU transmits on one of the top channels and do not collide with each other. $r^u_t$ is the reward at time $t$ for SU $u$, $I_t^u$ denotes the channel selected by SU $u$ at time $t$, $\mu_{I_t^u}$  and $C_{I_t^u}^u$ denote the vacancy probability and collision indicator on channel  $I_t^u$. If there is a collision, collision indicator is set to $1$. Otherwise, it is $0$. 
	 
The average number of collisions, $C_T$, is given by
\begin{equation}
\label{coll}
C_T=\sum_{t=1}^T\sum_{u=1}^U \mathbb{E}[C_{I_t^u}^u].
\end{equation}

		Our goal is to design distributed algorithms that keep regret and collisions as small as possible for static as well as dynamic networks. The various notations and their definitions are summarized in Table I.

	\renewcommand{\arraystretch}{1.2}
\begin{table}[!h]
	\captionsetup{justification=raggedright,singlelinecheck=false}
	\caption{Notations and Definitions}
	\label{gc}
	
	\begin{tabular}{|l|l|}
		\hline
		\textbf{Notations} & \textbf{Definitions} \\
		\hline
		$U$ & No. of SUs \\
		\hline
		$N$ & No. of channels\\
		\hline
		$t$ & Current time slot\\
		\hline
		$T$ & Length of the time horizon\\
		\hline
		$\mu_i^u$ & Vacancy probability of channel $i$ observed  \\ & by  the $u^{th}$ SU \\
		\hline
		$\Delta$ & Gap between $n^{th}$ and $(n+1)^{th}$ channel statistics \\
		\hline
		$R_{op}$ & Maximum total throughput achievable            		for SUs \\
		\hline
		$r_t^u$ & Reward received by the $u^{th}$ SU at time $t$\\
		
		\hline		
	 $I_t^u$ & Channel selected by the $u^{th}$ SU at time $t$ \\
	   \hline	
	   $C_{I_t^u}^u$ & Collision indicator for the $u^{th}$ SU at time $t$ for channel  $I_t^u$ \\ 
	   \hline
	   $C_T$ & Average number of collisions \\		
		
				\hline			
		$S_n^u$ & Number of times the channel $n$ is chosen by the $u^{th}$ SU \\
		\hline
		$V_n^u$  & Number of times the channel $n$, when chosen by the \\ & $u^{th}$ SU, is found vacant\\
		\hline
		$C_n^u$ &  Expected number of collisions seen by the $u^{th}$ SU  \\& over channel $n$ \\
		\hline
		$\pi^u$ & Array of the channel indexes sorted in the decreasing\\ &  order of estimated vacancy probabilities\\
		\hline
		$M_i $ & Number of times the channel $i$ needs to
be sensed \\ & to guarantee that the channel is found vacant at least once\\
%
\hline		
$T_{RH}$ & Duration of the random hopping phase\\
\hline		
$T_{SH}$ & Duration of the sequential hopping phase\\
\hline		
$T_{CC}$ & Duration of the CC phase\\
\hline
$T_{TR}$ & Total orthogonalization time of the trekking
phase\\
\hline
$T_{EN}^u$ & Time slot in which the $u^{th}$ SU enters into the network\\
\hline
$T_{BCI}$ & Duration of Best channel identification phase\\
\hline
$T_{TL}$ & Duration of the TL state \\
\hline
$N_m$ & Number of SUs who enter in the network at the start \\ & of the horizon\\
\hline
$N_e$ & Number of SUs who enter late in the network\\
\hline
$N_l$ & Number of SUs who leaves the network\\

		\hline
		$C_{T}$ & Average number of SU collisions in horizon \\
		\hline
		$R_{T}$ & Average regret \\
		\hline
		
	\end{tabular}
\end{table}	
\vspace{-0.2cm}
\section{OSA in Static Network}\label{SN}
In this section, we present proposed trekking based OSA algorithm for the static network (TSN) where the number of active SUs in the network is fixed but unknown. The TSN algorithm is run independently at each SU terminal. The algorithm has two phases namely, 1) Channel characterization (CC) phase and 2) Trekking phase. In the CC phase, the primary goal of each SU is to characterize the quality of channels via collision free hopping. In the trekking phase, SUs lock themselves in one of the top channels without the need of explicit $U$ estimation. The orthogonalization in the top channels guarantees zero regret thereafter.

 \subsection{TSN Algorithm}
 The TSN algorithm for a particular SU is given in Algorithm $1$. The same algorithm is run by all the SUs. The two subroutines, namely CC phase and Trekking phase are run sequentially. The CC phase is given in Subroutine $1$.  In this phase, each SU hops onto a channel selected uniformly at random (line $8$) in each time slot till it observes a collision-free transmission. Once such a channel is found (line $14$), the SU stops random hopping and starts sequential hopping (line $5$). In sequential hopping, a channel with higher index (up to modulo $N$) is selected in each slot (line $6$). The CC phase runs for $T_{CC} = (T_{RH}+T_{SH})$ number of time slots, where the values of $T_{RH}$ and $T_{SH}$ are specified in Lemma \ref{lma:RandomHop_Static} and \ref{lma:SeqHop_Static}. Note that $T_{RH}$ and $T_{SH}$ do not depend on $U$ and we use their lower bound while calculating $T_{CC}$ (line 2).  For clarity of notations, we omit the superscript $u$ in Algorithm $1$ and its subroutines.

\begin{algorithm}[!b]
	\caption{TSN Algorithm} \label{Algo1}
	\begin{algorithmic}
		\State	Input: {$N, \delta$ }
		\State	$(\pi,\{\hat{\mu}_i\})= CC(\delta, N)$ 
		\State	 Trekking$( \pi, \{\hat{\mu}_i\}, \delta)$ 
	\end{algorithmic}
\end{algorithm}

 \begin{figure*}[!b]
	\centering
	\subfloat[]{\includegraphics[scale=0.3]{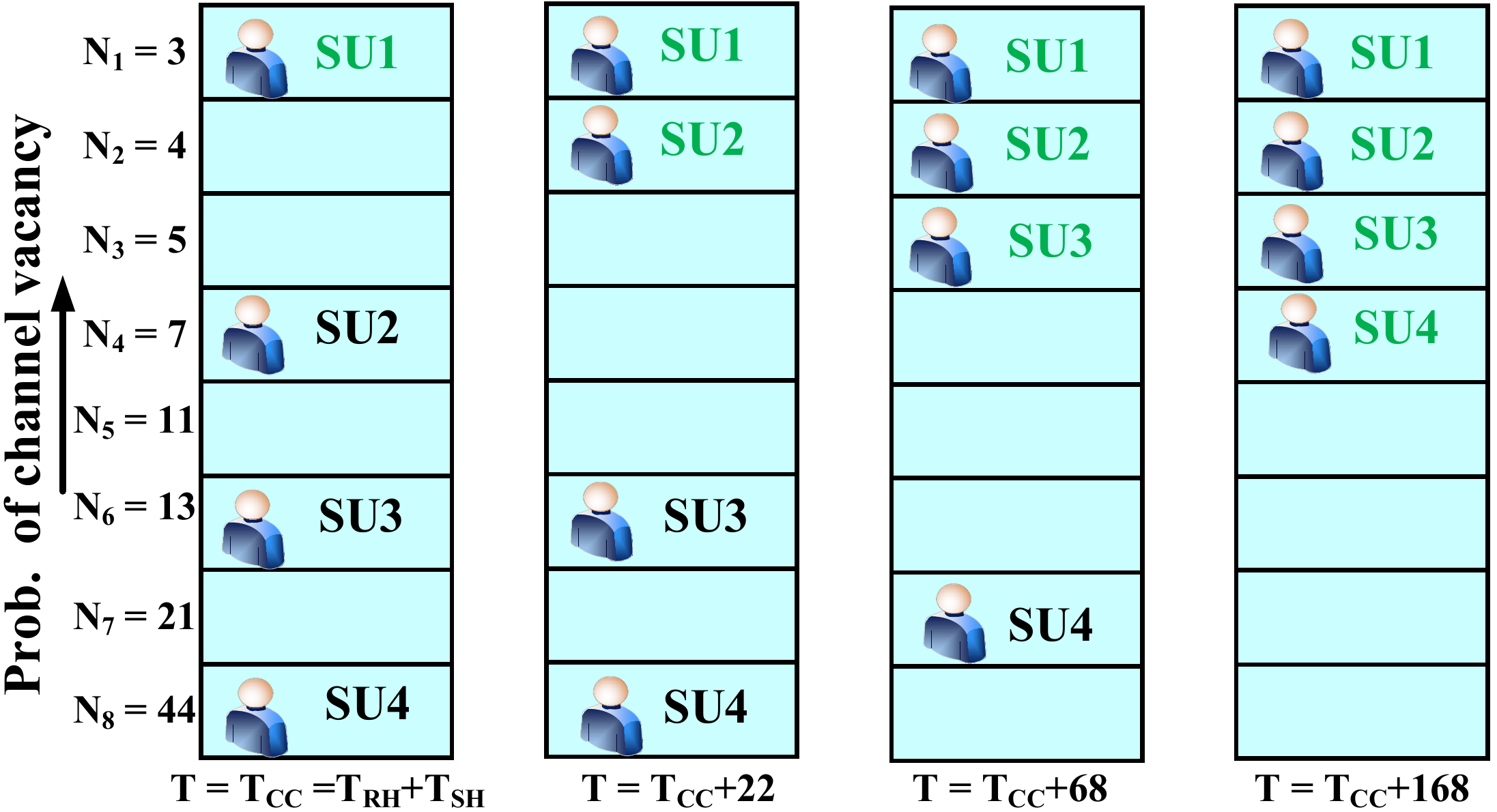}%
		\label{trek1}}
	\hspace{0.1cm}
	\subfloat[]{\includegraphics[scale=0.3]{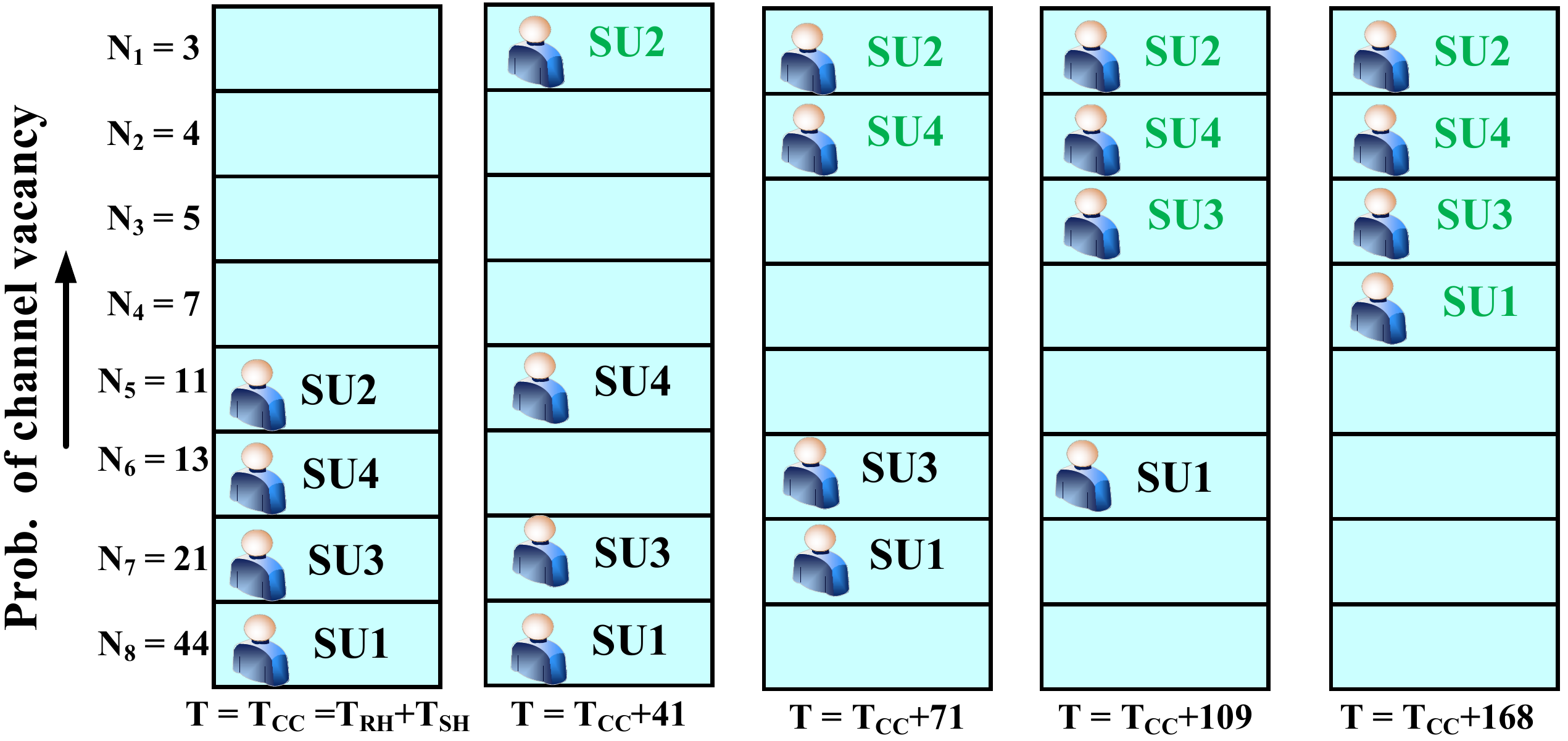}
		\label{trek2}}
	\vspace{-0.2cm}
	\caption{{\footnotesize Illustrative example for description of trekking phase of TSN algorithm for network with $N = 8$ and $\mu=[0.8,  0.7,... ,0.1]$ with (a) SU1, SU2, SU3 and SU4 in the channel with index 1, 4, 6 and 8, respectively at the end of CC phase, and (b) all SUs in bottom channels at the end of CC phase.} }
	\label{trek}
\end{figure*}

In each round, each SU senses the selected channel to check if it is vacant or occupied and transmits if it is vacant, otherwise it does not transmit.  Depending on the sensing and transmission feedback, each SU updates how many times each channel is selected ($S_n$ in line $10$) and how many times it is found vacant ($V_n$ in line $12$). At the end of the CC phase, the channel statistics $\hat{\mu}_i$ are estimated which are then used to rank the channels. The array $\pi$ contains the channel indices sorted in the decreasing order of the estimated vacancy probabilities. 

\begin{algorithm}[!t]
	\caption*{\textbf{Subroutine 1:} CC Phase of TSN}
	\begin{algorithmic}[1]
		\State Input: $N, \delta$
		\State Compute $T_{CC}=T_{RH}+T_{SH}$ using Eq.~\ref{eqn:TRH_Static} and Eq.~\ref{eqn:TSH_Static}
		\State Set $l=0$, $V_n=0,S_n=0 \;\; \forall n \in [N]$ and $r_t=0\;\; \forall t \in [T] $ 
		\For{$t=1 \dots T_{CC}$}
		\If{($l == 1$)}
		\State Choose channel, $I_t = I_{t-1}+1$ modulo $N$
		\Else
		\State Randomly choose channel, $I_t \sim U(1, ... ,N)$ 
		\EndIf
		\State Increment $S_{I_t}$ by 1
		\If{($I_t$ is vacant)}
		\State Increment $V_{I_t}$ by 1 and transmit over $I_t$
		\If{ (no collision)}
		\State Set $l=1$ and collect reward $r_t=1$
		\EndIf
		\EndIf	
		\EndFor	
		\State Estimate the channel statistics, $\hat{\mu}_n = \frac{V_{n}}{S_{n}} \;\; \forall n$
		\State Return set $\pi$ containing channel indices sorted according to decreasing values of $\hat{\mu}_n$
	\end{algorithmic}
\end{algorithm}

At the end of the CC phase, all the SUs are guaranteed to be in orthogonal channels with high probability. However, they need not be on the top channels. The objective of the second subroutine, i.e., trekking phase, is to move each SU to one of the top channels without estimating $U$. The pseudo code of trekking phase is given in Subroutine $2$. 

If an SU operates on a channel, say $i$, at the end of CC phase and finds that it is not the top-ranked channel (based on its channel estimates), then it aims to move to the next best channel provided it is not occupied by another SU. Otherwise it `falls-back' to channel $i$ and uses it until the end of time horizon (lines $16$-$17$). The channel $i$ is reserved for the SU while it checks for the availability of the next best channel and is released for other SUs only when SU vacates it. When an SU falls-back on its reserved channel, we refer to it as `locked' on that channel. Each SU keeps moving to the next best channel till they get locked on a channel. This process ensures that all the SUs orthogonalizes in the top channels without any communication or coordination among them.

The SU that has recently shifted to channel $i$ observes its next best channel for $M_i$ number of rounds (line $9$) to check if it occupied by another SU, where 
\[M_i=\sum_{j< i} N_j  \quad \mbox{and} \quad  N_j=\lceil \log (\delta/3)/(1-\hat{\mu}_j) \rceil.\]
Observing channel $k$ for $N_k$ time slots guarantees that, with probability at least $1-\delta/3$, the channel will be found vacant and hence the presence of an SU can be observed. Note that the SU on channel $i$ needs to observe its next best channel for $M_i$ slots and not just $N_i$ slots to check if it is occupied by another SU. This is because other SU would also be attempting to occupy their next best channel and waiting for $M_i$ slots ensures locking of all the SUs in channels better than channel $i$ and hence avoiding taking their reserved channels before they are released. The TR phase runs for at most $T_{TR}$ number of time slots, where the value of $T_{TR}$ is specified in Lemma \ref{lma:Trekking_Static}. Note that though $T_{RH}$ in Eq.~\ref{eqn:TTR_Static} depends on $U$ which is unknown, user do not need to calculate $T_{RH}$ to run trekking phase. It is only used to obtain regret bound in Theorem 1. To avoid collision among SUs before locking, SU follows long sensing before locking in the trekking phase where it first senses the presence of PU followed by the presence of SU. The long sensing is an efficient approach than collision since latter incurs reprocessing and retransmission penalty.

\begin{figure}[!h]
 	\centering
 	\subfloat[]{\includegraphics[scale=0.45]{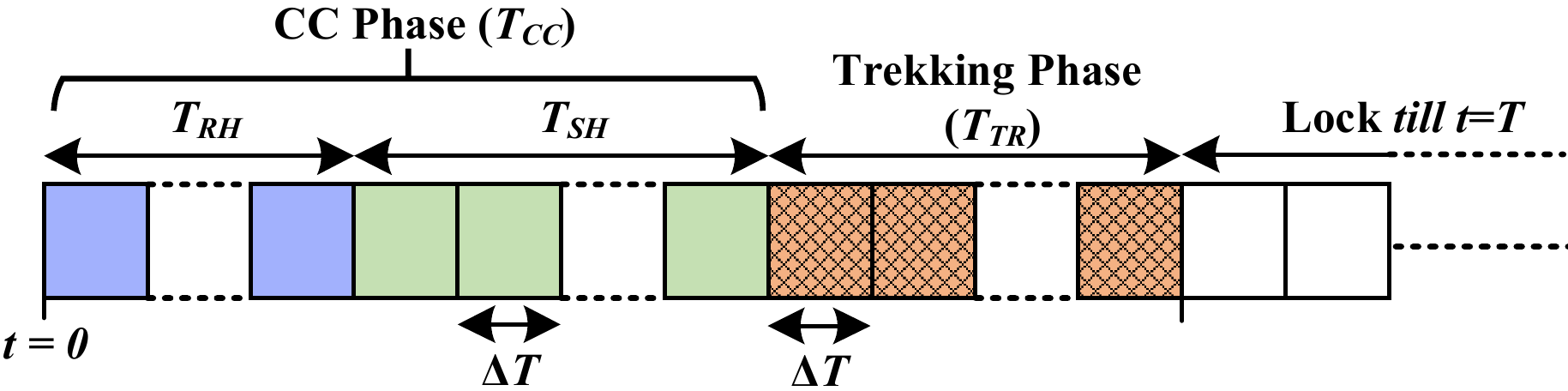}%
 		\label{p1}}\\ \vspace{-0.25cm}
 	\subfloat[]{\includegraphics[scale=0.37]{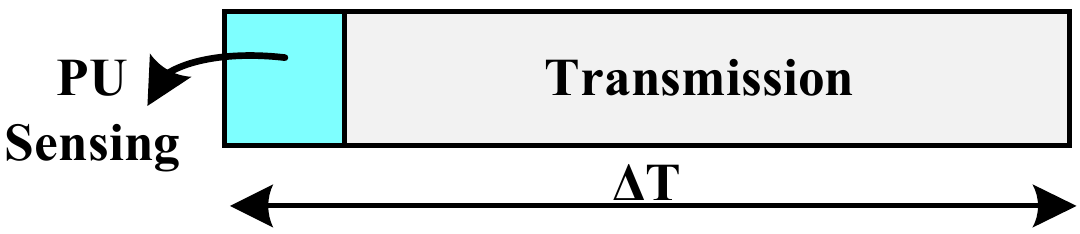}
 		\label{p2}}
 	\subfloat[]{\includegraphics[scale=0.37]{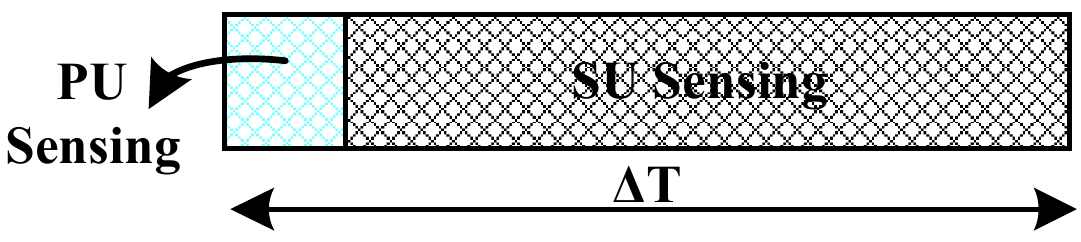}
 		\label{p3}}
 	\caption{{\footnotesize (a) Phases in the TSN algorithm at different instants of horizon, (b)~Sensing model for SUs in any phase except trekking phase, and (c) Sensing model in trekking phase.}}
 	\label{phase}
 	\vspace{-0.25cm}
 \end{figure}

The two phases or subroutines of the proposed TSN algorithm with respect to the horizon are shown in Fig. \ref{phase} (a). As discussed before, the time slot duration, $\varDelta T$, is same in each phase. The sensing model followed by SU during CC phase and after locking in the trekking phase involves only PU sensing as shown in Fig. \ref{phase} (b). In case of trekking phase, the unlocked SUs use long sensing model involving PU as well as SU sensing as shown in Fig. \ref{phase} (C).

\begin{algorithm}[!t]
	\caption*{\textbf{Subroutine 2:} Trekking Phase of TSN}
	\begin{algorithmic}[1]
		\State	Input : {$\pi, \{\hat{\mu}_i\}, \delta$ } 
		\State Re-index channels according to their rank in $\pi$
		\State Set $J$ to index of current channel of SU and $I_{T_{CC}}= J-1$
		\State Set $Y_i= 0\;\; \forall i \in [N]$ and  $L=0$ (channel lock indicator) 
		\State  Set $M_j =\sum_{i=1}^{j-1}\lceil \log (\delta/3) / (1-\hat{\mu}_i)\rceil \;\;\; \forall \;j$ 
		
		\For{$t=T_{CC}+1 \dots $}
		\If{$L==1$}
		\State Select the same channel $I_t=I_{t-1}$
		\ElsIf{$Y_{J}\leq M_{J}$}
		\State Select the same channel, $I_t=I_{t-1}$
		\Else
		\State Select the next best channel, $I_t=I_{t-1}-1$ 
		\State Set $J=I_{t-1}$ 
		\EndIf
		\State Increment $Y_{J}$ by 1
		\If{ $I_t$ is vacant and another SU present on it}
		\State Set $I_{t}=J$ and $L=1$	
		\EndIf
		\EndFor
	\end{algorithmic}
\end{algorithm}	
 
 For illustrations, consider the two scenarios shown in Fig. \ref{trek} with $N=8$ and $\mu=[0.8,  0.7,... ,0.1]$. The corresponding values of $N_i$ are shown for each channel. As expected, the value of $N_i$ increases as $\mu_i$ decreases. Consider Fig. \ref{trek}(a) where there are four SUs, SU1, SU2, SU3 and SU4 whose channel index at the end of CC phase are 1, 4, 6 and 8, respectively. After $M_4$=12 time slots, the SU2 in fourth best channel moves to the third best channel and in next $M_3$=7 time slots, i.e. in total 19 time slots, SU2 moves to the second best channel, i.e., $\pi_2$. After observing the top channel for $M_2=3$ time slots, the SU2 locks itself to the second best channel since SU1 is present in the top channel. At the same time, $SU3$ and $SU4$ are neither locked nor moved to next better channels due to higher values of $M_6(=30)$ and $M_8(=64)$.

   Likewise, after $M_6$=30 and $M_5$=19 time slots, SU3 moves to fifth and fourth best channel, respectively and finally after next $M_4$=12 time slots i.e. in total 61 time slots, SU3 moves to third best channels and locks itself in third best channel after $M_3=7$ time slots during which it senses the presence of SU2 on second best channel. Similarly, SU4 locks on the fourth best channel after 168 time slots. The TSN algorithm guarantees that the SU in the better channel at the end of CC phase locks before the SUs in any one of the worse channels. Similarly, SU3 moves to fourth best channel  after $M_8+M_7+M_6+M_5$=156 time slots and locks after $M_4=12$ more time slots. The TSN algorithm does not incur any regret thereafter. Note that SUs may transmit if they find idle channel during trekking phase (i.e. 168-time slots).

   The Fig. \ref{trek}(b) considers the worst case where the SUs are in bottom channels at the end of CC phase. Even then, the trekking approach guarantees that all SUs lock themselves in one of the top channels after 168 time slots. This means that the total duration of the trekking approach, $T_{TR}$, is independent of the number of SUs and the channels occupied by SUs at the end of CC phase and depends only on the channel statistics, $\mu_i$. Please refer to Lemma 3 for more details.

\subsection{Analysis of TSN Algorithm}
In this subsection, we bound the expected regret, and number of collisions of the TSN algorithm. We begin with the following definition given in \cite{MC}.

\begin{defn}
 An $\epsilon$-correct ranking of $N$ channels is a sorted list of empirical mean values of channel vacancy probabilities such that $\forall i, j: \mu_i$ is listed before $\mu_j$ if $\mu_i - \mu_j \geq \epsilon$.
\end{defn} 
The following theorems state a high confidence bound on
the expected regret and number of collisions of the TSN algorithm. The expectation is over the
algorithm's randomness.

 \begin{lemma}
 	\label{lma:RandomHop_Static}
 If each SU selects the channel uniformly randomly for $T_{RH}$ (See Eq. \ref{eqn:TRH_Static}) number of time slots, then all the SUs are on non-overlapping channels with probability at least $1-\delta_1$.
\end{lemma}

\noindent \textbf{Proof:} We want to compute $T_{RH}$ such that all the SUs are on non-overlapping channels with high probability within $T_{RH}$. If $P_C$ denote the collision probability of an SU when all the SUs are randomly hopping at any time slot $t$, and if none of the other SUs
are on the non-overlapping channel (worst-case) then the probability that the SUs will find
a non-overlapping channel within $T_{RH}$ is given by:
\begin{equation*}
\sum_{t=1}^{T_{RH}}P_C^{t-1}(1-P_C).
\vspace{-0.2cm}
\end{equation*}
We want this probability to be at least $1- \frac{\delta_1}{N}$ for each SU. Hence we set
\begin{eqnarray}
\label{trh}
\lefteqn{\sum_{t=1}^{T_{RH}}P_C^{t-1}(1-P_C) \geq 1- \frac{\delta_1}{N}} \nonumber  \\
&\iff&1-{P_C}^{T_{RH}} \geq 1- \frac{\delta_1}{N} \nonumber  \\
&\iff& T_{RH} \log{P_C} \leq \log\bigg({\frac{\delta_1}{N}}\bigg) \nonumber  \\
&\iff& T_{RH} \geq \frac{\log\big({\frac{\delta_1}{N}}\big)}{\log{P_C}}.
\vspace{-0.5cm}
\end{eqnarray}
To obtain $T_{RH}$, we next bound $P_C$. Let $p_{ns}$ denote the probability of no collision due to non-settled SUs (i.e. RH SUs) and
$p_{s}$ denote the probability of no collision due to settled SUs (i.e. SH SUs). We have
\begin{align}
P_{NC}&= 1- P_C
= \sum\limits_{i = 1}^{N}\frac{\mu_i}{N}\left(p_{ns}+p_{s}\right) + \sum\limits_{i=1}^{N}\frac{(1-\mu_i)}{N} \nonumber \\	
&\geq \sum\limits_{i = 1}^{N}\frac{\mu_i}{N}\left(p_{ns}+p_{s}\right) \geq \sum\limits_{i = 1}^{N}\frac{\mu_i}{N}p_{ns} \nonumber\\& \geq \sum\limits_{i = 1}^{N}\frac{\mu_i}{N}\left(1-\frac{1}{N} \right)^{U-1}  \geq \left(1-\frac{1}{N} \right)^{U-1}\theta \nonumber \\ &  > \left(1-\frac{1}{N} \right)^{N-1}\theta.
\end{align}
where we used the relation $\sum\limits_{i = 1}^{N} \mu_i/N > \theta$ in the second last inequality. Substituting the bound on $P_C$ in Eq.~\ref{trh}, we get
\begin{equation}
\label{eqn:TRH_Static}
T_{RH} \geq \frac{\log \left(\frac{\delta_1}{N}\right)}{\log \left(1-\theta\left(1-\frac{1}{N} \right) ^ {N-1} \right)}
\end{equation}
\hfill\IEEEQED
%

 \begin{lemma}
 	\label{lma:SeqHop_Static}
 After initial $T_{RH}$ time slots, if each SU selects the distinct channel via sequential hopping for $T_{SH}$ (See Eq. \ref{eqn:TSH_Static}) number of time slots, then with probability at least $1-\delta_2$ all the SUs will have $\epsilon$-correct $(\forall \epsilon >0)$ ranking of channels.
 \end{lemma}
\noindent \textbf{Proof:} \textbf{Channel Ranking Estimation:} If for any SU $u$ it is true that $\forall n \in 1 \cdots N$ $\left | \hat{\mu}_n-\mu_n \right| \leq\frac{\epsilon}{2} $, then the 
player has an  $\epsilon-$ correct ranking. We will upper bound the probability that no SU has $\epsilon-$correct ranking given the SU have $O_{min}$ observations of each 
channel. 
Consider the following events: \\

\noindent$J_u$ - event that a SU $u$ has observed each channel at least $O_{\min}$ number of times.\\
$A$ - event that all SUs have an $\epsilon$-correct ranking. \\
$A_u$ - event that a SU $u$ has $\epsilon$- correct ranking. \\
$B$ - event that all SUs have atleast $O_{min}$ observations of each channel. \\
$B_u$ - event that a SU $u$ has atleast $O_{min}$ observations of each channel. \\ 

We want to compute,
\[Pr(\overline{A}_u|B_u) < \frac{\delta_2}{N} \]

Note $\overline{X}$ denotes complement of any event $X$. Then,
\begin{align*}
Pr(\overline{A}_u|B_u) &\leq Pr\left ( \exists n \in 1 \cdots N \ \ s.t | \hat{\mu}_n-\mu_n | >\frac{\epsilon}{2} | \ B_u\right) \\
 \tag{By Union Bound}      &\leq \sum_{n=1}^{N} Pr\left ( | \hat{\mu}_n-\mu_n | >\frac{\epsilon}{2} | \ B_u\right) \\ 
       			&= \sum_{n=1}^{N} \sum_{j=O_{min}}^{\infty} Pr\left (| \hat{\mu}_n-\mu_n | >\frac{\epsilon}{2} | \ 
J_u = j \right) \cdot \\
& \hspace{2.1cm} Pr \left( J_u = j| \ B_u \right) \\
 \tag {By Hoeffding's Inequality} &\leq \sum_{n=1}^{N} \sum_{j=O_{min}}^{\infty} 2 \cdot \exp \left(\frac{-j \cdot \epsilon^2}{2} \right) Pr \left( J_u = j| \ B_u \right)  	
\end{align*}
 \begin{align*}
 &\leq \sum_{n=1}^{N} 2 \cdot \exp \left(\frac{-O_{min} \cdot \epsilon^2}{2} \right) \sum_{j=C}^{\infty} Pr \left( J_u = j| \ B_u \right) \\
   &\leq \sum_{n=1}^{N} 2 \cdot \exp \left(\frac{-O_{min} \cdot \epsilon^2}{2} \right) \\
		&\leq N \cdot 2 \cdot \exp \left(\frac{-O_{min} \cdot \epsilon^2}{2} \right)
\end{align*}

We can apply Hoeffding's Inequality since each observation of the channel is independent of the number of times we observe that channel. In order for this to be $< \frac{\delta_2}{N}$, 
	 \[ N \cdot 2 \cdot \exp \left(\frac{-O_{min} \cdot \epsilon^2}{2} \right) < \frac{\delta_2}{N} \newline
		\implies  O_{min} >  \frac{2}{\epsilon^2} \cdot \ln \left ( \frac{2 \cdot N^2}{\delta_2} \right) 
\]

\noindent We note that each SU gets one observation of a channel in each time slot due to collision-free sequential hopping. Thus, the number of time slots required to obtain $O_{min}$ observations of all the channels, i.e., $T_{SH}$, is given by:

\begin{equation}
\label{eqn:TSH_Static}
T_{SH} \geq \frac {2 \cdot N}{\epsilon ^ 2} \cdot \ln\bigg(\frac{2 \cdot N^2}{\delta_2}\bigg)
\end{equation}


\hfill\IEEEQED

\begin{lemma}
 \label{lma:Trekking_Static}
 In $T_{TR}$ (See Eq. \ref{eqn:TTR_Static}) time slots of trekking phase, all the SUs will settle in one of the top channels with probability at least $1-\delta_3$.\end{lemma}
 
 \noindent \textbf{Proof:} The number of time slots required for any SU recently shifted to channel $i$ to observe its next best channel to check if it is occupied by another SU, denoted as $M_{i}$, is given by:

\begin{equation}
\label{c}
M_i=\sum_{j< i} N_j
\end{equation}
where $N_{j}$ is the number of time slots required to guarantee that, with
probability at least $1-\delta^\prime$, the channel will be found vacant
and hence presence of the SU can be observed. It is given by:
\begin{equation}
\label{z}
\sum\limits_{t = 1}^{N_{j}} (1-\mu_j)^{t-1}(\mu_j)\geq 1-\delta^\prime
\end{equation}
\begin{equation}
N_{j} \geq \bigg\lceil\frac{\log\delta^\prime}{\log{(1-\mu_j)}}\bigg\rceil
\end{equation}
where $\mu_j$ is the vacancy probability of the $j^{th}$ channel. The first term in Eq.~\ref{z} is the probability that $j^{th}$ channel is occupied till the time slot $t-1$ and is vacant in the $t^{th}$ time slot. An upper bound on the total time slots, $T_{TR}$, required by all the SUs to settle in one of the top channels can be obtained as
\begin{eqnarray*}
\label{l3}
T_{TR} &=& \sum\limits_{i = 2}^{N} M_i= \sum\limits_{i = 2}^{N}  \sum\limits_{j = 1}^{i-1} \bigg\lceil\frac{\log\delta^\prime}{\log{(1-\mu_j)}}\bigg\rceil \\
 \hspace{-.2cm} &\geq& \sum\limits_{i = 2}^{N}  \sum\limits_{j = 1}^{i-1} \bigg\lceil\frac{\log\delta^\prime}{\log{(1-\theta)}}\bigg\rceil \geq \bigg\lceil\frac{\log\delta^\prime}{\log{(1-\theta)}}\bigg\rceil \frac{(N-1)(N)}{2}, 
\end{eqnarray*}
where we used the the relation $\mu_j > \theta$ in the first inequality. By setting $\delta^\prime=1-\delta_3/(KN)$, it is guaranteed that each player gets the correct observation on each channel with probability at least $\delta/3.$ Thus,
 
\begin{equation}
\label{eqn:TTR_Static}
T_{TR} = \bigg\lceil\frac{\log(\delta_3/NU)}{\log{(1-\theta)}}\bigg\rceil \frac{(N-1)(N)}{2}.
\end{equation}\hfill\IEEEQED
\begin{thm}
	\label{thm:Regret_Static}
For all $\delta \in (0,1)$, with probability  $\geq 1-\delta$, the expected regret of the network consisting of $U$ SUs running the TSN algorithm with $N$ channels for horizon of size $T$ is upper bounded by:  $R_T\leq U[T_{RH}+T_{SH} \cdot (1 - \frac {U}{N}) + T_{TR}]$, where the value of $T_{RH}$, $T_{SH}$ and $T_{TR}$ are given in Eq. (\ref{eqn:TRH_Static}), (\ref{eqn:TSH_Static}) and (\ref{eqn:TTR_Static}), respectively. 

\end{thm} 

\noindent The first and second term is due to the regret incurred by the SUs in the CC phase which runs for $T_{CC} = T_{RH}+T_{SH}$ number of time slots. The third term corresponds to regret incurred in the $T_{TR}$ duration. For $t >T_{CC}+ T_{TR}$, the regret is zero since all SUs are orthogonalized on the top channels. \\

 \noindent \textbf{Proof:} Let $Y$ denote the intersection of the following three events:
\begin{itemize}
	\item Event $A$--all SUs are orthogonalized after $T_{RH}$ number of slots 
	\item Event $B$--all SUs have the correct ranking of channels after $T_{SH}$ time slots 
	\item Event $C$-- all SUs are settled in one of the top channels in $T_{TR}$ number of slots
\end{itemize}


Using Lemmas $(\ref{lma:RandomHop_Static}),( \ref{lma:SeqHop_Static}),(\ref{lma:Trekking_Static})$, the event $Y$ holds with probability at least 
\begin{equation*}
Pr(Y)  = Pr(A) Pr(B|A )Pr(C|B,A)
\end{equation*}

\begin{equation*}
\geq (1-\delta/3)^3 \geq 1-3\frac{\delta}{3} \geq 1-\delta.
\end{equation*}

Setting  $\delta_1 = \delta_2=\delta_3=\frac{\delta}{3}$

\begin{equation*}
 \geq (1-\delta/3)^3 \geq 1-3\frac{\delta}{3} \geq 1-\delta
\end{equation*}


%
%

For $t > T_{RH}+T_{SH}+T_{TR}$, all the SUs are orthogonalized on the top channels, hence regret is zero with probability at least $1-\delta$. For any $t \leq T_{RH}+T_{SH}+T_{TR}$, the regret due to each SU can be upper bounded by $t$ since regret per SU per time slot is at most $1$. Hence total regret is bounded by 
\[ R_T \leq U  \left[T_{RH}+T_{SH}+T_{TR} \right]  \] with probability at least $1-\delta$. Note that we use lower bounds on $T_{RH}$, $T_{SH}$, and $T_{TR}$ obtained using above lemmas.
Now notice that during the $T_{SH}$ duration, each SU spends a fraction $U/N$ of the time slots on one of the top channels. Compared to an optimal allocation of the channels to the SUs where they operate on one of the top channels throughout,  this results in zero regret from the SUs during this fraction of the time slots. Hence the above upper bounds can be tightened as 
 \[ R_T \leq U  \left[T_{RH}+T_{SH}(1-U/N)+T_{TR} \right] . \]
 This concludes the proof of Thm \ref{thm:Regret_Static}. 
 \hfill\IEEEQED

\begin{remark}
As the number of SUs in the network increases to the number of channels, i.e., $U = N$, every SU will be selecting one of the top channels in SH phase. This leads to zero regret in SH phase and hence, the expected regret of the network will be $R_{T} \leq N \cdot \big[ T_{RH} + M_N \big]$.
\end{remark}
Thus, as the number of SUs in the network increases, the regret of the TSN algorithm decreases as opposed to existing state-of-the-art algorithms whose regret increases with increase in $U$. We also validate this via simulation and experimental results in Section ~\ref{sim_res} and \ref{exp_res}, respectively.  

\begin{remark}
	Note that the above regret bounds are pessimistic as they did not account for the fact that channels are busy for at most $(1-\theta)$ fraction of the time slots. One should be able to improve the above regret bounds during the sequential hopping and the trekking phase by a factor of at least $1- \theta$. 
\end{remark}

\begin{thm}
	\label{thm:Collsions_Static}
For all $\delta \in (0,1)$, with probability  $\geq 1-\delta$,  the expected number of SU collisions in the TSN algorithm with $N$ channels for horizon of size $T$ is upper bounded by  $U \cdot T_{RH}$. \\
 For $t > T_{RH}$, the number of collisions is zero since all SUs are orthogonalized in different channels. \\
\end{thm} 

\noindent \textbf{Proof:} Here we upper bound the number of collisions incurred in the TSN algorithm. In the CC phase, collisions take place only during $T_{RH}$ due to the random selection of channels whereas there is no collision in the $T_{SH}$ duration due to orthogonalized sequential hopping by the SUs. Also in trekking phase, there is no collision among the SUs due to the long sensing by the SUs in the $T_{TR}$ duration and zero collision afterwards due to locking of the SUs in the top $U$ channels. Thus, the number of collisions incurred
in the TSN algorithm with $N$ channels for $T$ rounds is upper bounded by $U \cdot T_{RH}$ with probability  $\geq 1-\delta$. \hfill\IEEEQED

%
%
%
%
%
%
%

\vspace{-0.2cm}
\section{OSA IN DYNAMIC NETWORK}\label{DN}
In this section, we adopt TSN algorithm to more challenging dynamic networks where SUs can enter and leave the network anytime. The proposed algorithm is referred to as Trekking in Dynamic Networks (TDN). It consists of two phases namely, 1) Channel characterization (CC) phase and 2) Continuous Trekking (CTR) phase that run sequentially.  Since the SUs in the dynamic case can enter or leave the network anytime, some of the SUs will be in the CC phase while the others in the CTR phase at any given time. 
While both CC phase and CTR phase are similar to the CC phase and the Trekking phase of the TSN in functionality, both phases have to deal with new SUs joining and leaving the system dynamically. Dynamic users can be easily dealt in the CC phase using the long sensing, but it is more challenging in the CTR phase as any top channel vacated by SUs leaving the system should be taken over by existing SUs and hence no existing SU (unless already in on the top channel) should lock on any channel permanently.  The word `continuous' in the CTR phase signifies this act of continuously trekking towards the top channel without locking to any channel permanently till they leave.

 \subsection{TDN Algorithm}
 
 The proposed TDN  is given in Algorithm $2$. Whenever the SU enters into the network, it is in CC phase by default. Similar to TSN algorithm, CC phase consists of random hopping and sequential hopping for $T_{CC}=T_{RH} +T_{SH}$ time slots, respectively. The major difference with respect to TSN algorithm is that SU needs to distinguish between the SUs that are in CC phase and the CTR phase. To do this, SUs entering the system looks for presence of the SU on a channel using the long sensing model (Fig.~\ref{p3}) instead of short sensing model (Fig.~\ref{p2}) in the CC phase of the TSN algorithm so that they do not collide with the existing SUs. At the end of CC phase, each SU calculates the channel raking based on the estimated channel statistics. The subroutine for CC phase of TDN is exactly similar to CC phase of TSN algorithm and is omitted.

\begin{algorithm}[!h]
	\caption{TDN Algorithm} \label{Algo2}
	\begin{algorithmic}
		\State Input: $ T_{TL}, N, \delta$ 
		\State $(\pi,\{\hat{\mu}_i\})= CC( N, \delta)$ 
		\State $ CTR (\pi, \{\hat{\mu}_i\},T_{TL},\delta)$
	\end{algorithmic}
\end{algorithm}

After the CC phase, each SU enters into the CTR phase immediately. The pseudo code for the CTR phase is given in Subroutine $3$. Similar to the trekking phase of TSN algorithm, the aim of the CTR phase is to move the SUs to one of the available top channels. Note that in a dynamic network, previously occupied channel can become unoccupied in the future and hence the SUs should periodically look for the availability of top channels. To account for this, we introduce two states for SUs in the CTR phase namely : 1) Temporary Locking (TL) state, and 2) Best Channel Identification (BCI) state between which the SUs alternate.  When an SU starts the CTR phase on a channel, say $i$, its default state is BCI. Since this channel could be occupied by another SU, the SU first observes it for $M_{i-1}$ number of slots (line $8$) and if it is found to be unoccupied, then the SU occupies it and enters into the TL state. Otherwise, it checks for the availability of channel $i+1$ (next worst to channel $i$) and observes it for $M_i$ number of slots. The SU repeats this process (trekking downwards) till it enters the TL state for the first time. Once the SU enters the TL state on a channel, it locks itself on that channel for $T_{TL}$  number of slots before it starts checking for availability of next best channel, i.e., trekking upwards (line $24$). When the SU finds a non occupied channel for the first time, the parameters FB is set to $1$  (line $20$). Once FB is set to $1$, the SU treks upwards as it is now guaranteed to have one non-occupied or reserved channel that it can 'fall-back' in case the better channels are occupied.

\begin{algorithm}[!t]
	\caption*{\textbf{Subroutine 3:} CTR phase of TDN }
	\begin{algorithmic}[1]
		\State	Input : {$\pi, \{\hat{\mu}_i\},  T_{TL},\delta $ } 
		\State Re-index channels according to their rank in $\pi$
		\State Set $I_{T_{CC}}$ = index of the SU channel and $J=I_{T_{CC}}+1$
		\State Set $Y_i= 0\;\; \forall i \in [N]$ and  $X=0, TL=0, FB=0$ 
		\State  Set $N_i=  \lceil \log (\delta/3) / (1-\hat{\mu}_i)\rceil$ and $M_j =\sum_{i=1}^{j-1}N_i$
		\For {$t=T_{CC}+1 \dots $}	
		\If{$TL==0$}
		\If{$Y_{J}<M_{J}$ }
		\State Select the same channel, $I_t=I_{t-1}$
		\State Increment $Y_J$ by 1
		\If{ another SU present on $I_t$}
		\If{ $FB==1$}
		\State Set $I_t=J, TL=1,X=0$
		\Else
		\State Set $I_t=J, J=J+1, Y_J=0$
		\EndIf 
		\EndIf
		\Else
		\State Select the next best channel, $I_t=I_{t-1}-1$, 
		\State Set $J=I_{t-1},Y_J=0, FB=1$
		\EndIf
		\Else
		\If{$X \leq T_{TL}$}
		\State Select the same channel $I_t=I_{t-1}$
		\State Increment $X$ by 1
		\Else
		\State $TL=0, I_t=I_{t-1}-1, J=I_{t-1}, Y_J=0$
		\EndIf	
		\EndIf            
		\EndFor
	\end{algorithmic}
\end{algorithm}	

Note that, after entering into the CTR phase, each SU treks downwards till the parameters FB is set to $1$ and then continues to trek upwards. CTR phase is designed so because when an SU starts the CTR phase, it is likely that the top channels are taken by the SUs that entered before it. Hence it should check for availability of a lower ranked channel, and once a non-occupied channel is found, it can check for better channels that are vacated by the leaving SUs. 

In the TL state, SU follows short sensing approach and use the same channel for $T_{TL}$ rounds. The value of $T_{TL}$ is an input parameter that should be specified based on the rate of the leaving of the SUs. Typically its value should be set small if the rate of leaving SUs is high. Otherwise, it should be set high. When an SU enters the TL state on a channel, that channel is reserved for the SU. This is guaranteed by the fact that each SU observers channel $i$ for $M_i$ slots before occupying it (see discussion in the TSN algorithms).

\par

	\begin{figure*}[!t]
	\centering
	\includegraphics[scale=0.55]{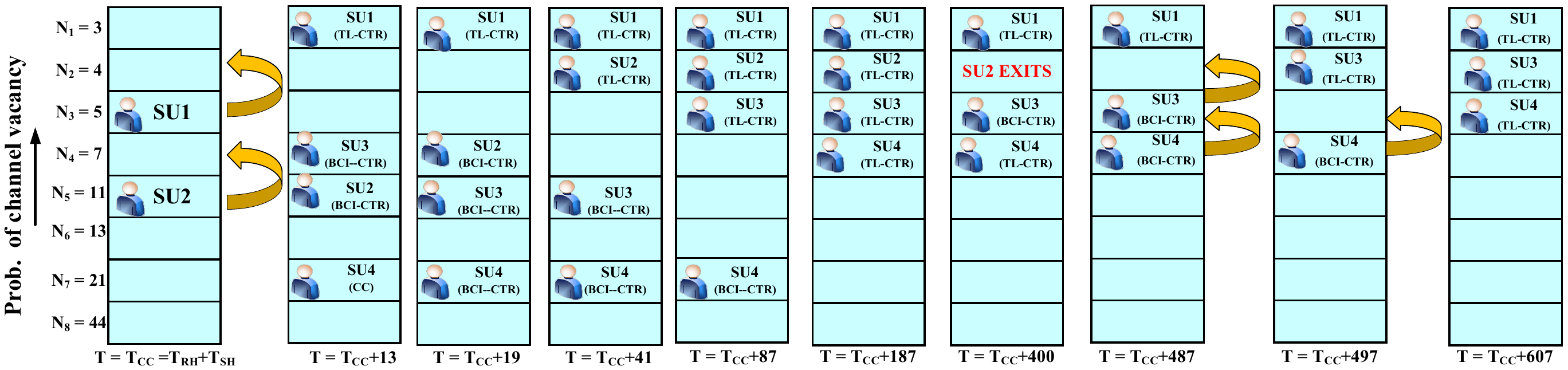}%
	\caption{{\footnotesize Illustrative example for description of CTR phase of TDN algorithm for network with $N = 8$ and $\mu=[0.8,  0.7,... ,0.1]$.}} 
	\label{trekking2}
\end{figure*}

 For illustrations, consider the scenario shown in Fig. \ref{trekking2} with $N=8$ and $\mu=[0.8,  0.7,... ,0.1]$. The corresponding values of $N_i$ are shown for each channel. In the beginning, two SUs, SU1 and SU2, enter the network at the start of a horizon and hence, they are in the orthogonal channels at the end of their CC phase. For instance, SU1 and SU2 are in the channel with index 3 and 5, respectively at the end of CC phase. After $M_3+M_2+M_1$=13 time slots, the SU1 locks itself in the best channel, i.e. $\pi_1$. At the same time, a new SU, SU3, who have entered late in the network completes its CC phase and reaches at the channel with index 4, but SU2 remains in the fifth channel due to higher $M_5$ value. The SU4 is still in its CC phase at this time. The SU3 faces the collision with SU2 who is still in its BCI state of CTR phase and thus SU3 hops to the fifth best channel and SU2 shifts to the fourth best channel within $M_5$=19 time slots. At this instant, another SU, SU4, reaches the channel with index 7 after completing its CC phase. 
 
 After $M_4+M_3+M_2$=22 time slots, i.e., in total 41 time slots, SU2 settles in the second best channel whereas SU3 is in BCI phase and still sensing the vacancy of the fifth best channel. After sensing the vacancy of the fifth best channel for $M_6$=30 time slots from its arrival on that channel, i.e., at $T_{CC}+19+30=T_{CC}+49$, SU3 confirms the vacancy of this channel and starts trekking to find the best available channel. After $M_5+M_4+M_3=19+12+7=38$ time slots, i.e., at $T_{CC}+49+38=T_{CC}+87$, SU3 settles on the third best channel whereas SU4 is still sensing the vacancy status of the channel with index 7. After sensing the vacancy status of the seventh best channel for $M_8$=64 time slots from its arrival on that channel, i.e., at $T_{CC}+19+64=T_{CC}+83$, SU4 confirms the vacancy of this channel and starts trekking to find the best available channel. After $M_7+M_6+M_5+M_4=43+30+19+12=104$ time slots, i.e. at at $T_{CC}+83+104=T_{CC}+187$, SU4 settles on the fourth best channel.
  
 Next, consider that the SU2 leaves the network at $T_{CC}$ + 400 time slot. Since every SU switches back and forth between TL and BCI state at a fixed interval of $T_{TL}$ (say 200) time slots, SU3 senses the next best i.e. $2^{nd}$ channel at $T_{CC}+87+200+200$ i.e. $T_{CC}+487$ time slot and after $M_3+M_2$=10 i.e. at $T_{CC}+497$ time slot, SU3 moves to the $2^{nd}$ best channel while SU1 remains in the top most channel. Similarly, SU4 moves to the third best channel at  $T_{CC}+187+200+200+M_4+M_3$ i.e. $T_{CC}+607$ time slot. Note that SUs can transmit if they find idle channel during any time slot of the CTR phase irrespective of the state of SU.

\subsection{Analysis of TDN Algorithm}
In this subsection we bound the expected regret of TDN .
\begin{lemma}
 \label{lma:TBCI_dynamic}
 In $T_{BCI}$ (See Eq. \ref{eqn:TBCI_dynamic}) time slots of BCI phase, all the SUs will settle in one of the top channels with probability at least $1-\delta_3$ ($0<\delta_3<1$).\end{lemma}

\noindent \textbf{Proof:} Observing a channel by the SUs entered in the BCI state of the CTR phase to confirm its availability in the TDN algorithm is same as observing the next best channel to check if it
occupied by another SU in the trekking phase of TSN algorithm, thus similar to $T_{TR}$ of TSN algorithm, $T_{BCI }$ can be given as:
\begin{equation}
\label{eqn:TBCI_dynamic}
T_{BCI} = \bigg\lceil\frac{\log\delta/3}{\log{(1-\theta)}}\bigg\rceil \frac{(N-1)N}{2}.
\end{equation}
\hfill\IEEEQED
\begin{thm}
\label{thm:Regret_dynamic}
Let $N_m$ be the number of SUs entering at the beginning and let $N_e$ and $ N_l$ be the total number of SUs entering and leaving the network, respectively, over time period $T$. Let $T_{BCI}$ be the duration of BCI phase. Then, for all $\delta \in (0,1)$, with probability at least $1-\delta$, the expected regret of TDN is upper bounded as \\
\begin{eqnarray*}
R_T \leq \bigg[ N_m [T_{RH}+T_{SH}  (1 - \frac {N_m}{N}) +T_{BCI}+x_0 M_{N}]
\end{eqnarray*} 
\begin{eqnarray}
	+ \sum_{i=1}^{N_e} [T_{RH}+T_{SH}+ T_{BCI}+ x_i M_{N}] + N_l  T_{TL} \bigg],
\end{eqnarray} 
where 
\begin{equation}
\label{eqn:TRH_dynamic}
x_0 = \bigg\lceil\frac {T - T_{CC} - T_{BCI}}{T_{TL} + T_{BCI}}\bigg\rceil
\end{equation}
\begin{equation}
\label{eqn:TRH_dynamic1}
x_i =\bigg\lceil \frac {T - T_{CC} - T_{BCI}-T_{EN}^i}{T_{TL} + T_{BCI}}\bigg\rceil \leq x_0
\end{equation}

\end{thm}
\noindent 
The value of $T_{RH}$, $T_{SH}$ and $T_{BCI}$ are given in Eqs (\ref{eqn:TRH_Static}), (\ref{eqn:TSH_Static}) and  (\ref{eqn:TBCI_dynamic}), respectively.
The first part in the bracket is the regret caused by the SUs that entered at the start of horizon. These SUs incur regret during the $T_{CC}$ and $T_{TR}$ duration similar to the TSN algorithm. In addition, the regret during $T_{BCI}$ duration is due to the maximum time taken by the SU to move to top channel from the worst channel. The $T_{BCI}$ is also refer to the maximum number of time slots a SU requires to enter into the TL state from the BCI state. The term $x_0\cdot T_{BCI}$ is the regret due to periodical BCI state during the horizon $T$. The second part corresponds to the regret due to the SUs who enter late in the network, i.e. for SU $i$ entering at time $T_{EN}^i$ and is identical to first part except smaller $X_i$. The third part  corresponds to the regret incurred due to the SUs leaving the network. Here $T_{TL}$ implies the worst case scenario in which the SU at the top most channel leave the network just after the time slot in which the SU at lower most channel get locked and thus it checks its next better channel after $T_{TL}$ time slots.\\ 

\noindent \textbf{Proof:} Let $N_m$ be the number of SUs who enter
in the network at the start of the horizon, $N_e$ and $N_l$ be
the total number of SUs entering and leaving the network. we compute a bound
on the expected regret of $U$ SUs running the TDN with $N$ channels for $T$ rounds. The regret is composed of three terms:

\begin{itemize}
\item
 Regret due to the SUs entered at the start of horizon 
 \item
 Regret due to the entering SUs
 \item
 Regret due to the leaving SUs
\end{itemize}

We will now compute each of these terms. \\
\textbf{Regret due to the SUs entered at the start of horizon:} We have at most $N_m$ SUs who enter at the start of horizon and incur regret in the CC and CTR phase. In the CC phase, they incur regret in the $T_{RH}$ and $T_{SH}$ duration except the $T_{SH} \cdot (\frac{N_m}{N})$ time slots during which the SUs select one of the \textquoteleft top\textquoteright $U$ channels without causing any regret.  Thus, the upper bound on the regret in CC phase is given by: 
\begin{equation}
\leq  N_m \cdot (T_{RH} + T_{SH} \cdot (1- \frac{N_m}{N}))
\end{equation}
In the CTR phase, SUs will switch between the BCI  
and TL state for checking the availability of next best channel. Thus, the number of times the SUs will switch between these two states is given by: 
\begin{equation}
x_0 = \bigg\lceil\frac {T - T_{CC} - T_{BCI}}{T_{TL} + T_{BCI}}\bigg\rceil
\end{equation}
where the subscript 0 indicates the user entering at the start of horizon. Thus the expected regret in the CTR phase is upper bounded by: 
\begin{equation}
\leq  N_m \cdot (T_{BCI}+ x_0 \cdot M_{N})
\end{equation}

 Thus, the expected regret due to the SUs entered at the start of horizon is upper bounded by: 
 \begin{equation}
\leq   N_m \cdot[T_{RH}+T_{SH} \cdot (1 - \frac {N_m}{N}) + T_{BCI}+ x_0 \cdot M_{N}]
\end{equation}

\textbf{Regret due to the entering SUs:}  
Unlike the SUs entered at the start of horizon, the newly entered SUs  may incur regret throughout the $T_{SH}$ duration as they will not be necessarily in one of the top channels in the $T_{SH} \cdot (\frac{U}{N})$ time slots. Thus, the regret incurred by the $N_e$ SUs in the CC phase is upper bounded by:
\begin{equation}
\leq  N_e \cdot (T_{RH} + T_{SH})
\end{equation}

The expected regret due to the entering SUs in the CTR phase is upper bounded by: 
\begin{equation}
\leq  \sum_{i=1}^{N_e} (x_i \cdot M_{N}+T_{BCI})
\end{equation}
where $x_i$ is the number of times the SU entering the network at time $T_{EN}^i$ will switch between BCI and TL states. It is given by,
\begin{equation}
\label{eqn:TRH_dynamic13}
x_i = \bigg\lceil\frac {T - T_{CC} - T_{BCI}-T_{EN}^i}{T_{TL} + T_{BCI}}\bigg\rceil \leq x_0
\end{equation}
Thus, the expected regret due to the entering SUs is upper bounded by:
\begin{equation}
\leq   \sum_{i=1}^{N_e} \cdot[T_{RH}+T_{SH} +x_i \cdot M_{N}+T_{BCI}] 
\end{equation} \\
\textbf{Regret due to the leaving SUs:} We assume the worst case scenario in which the SU at the top most channel leave the network just after the time slot in which the SU at lower most channel get locked and thus it checks its next better channel after $T_{TL}$ time slots. Thus, the expected regret due to the leaving SUs is upper bounded by: 
 \begin{equation}
\leq   N_l \cdot T_{TL}
\end{equation} 
Thus, the expected regret of $U$ SUs running the TDN with $N$ channels for $T$ rounds is upper bounded by:

 \begin{equation*}
  \bigg[ N_m \cdot[T_{RH}+T_{SH} \cdot (1- \frac{U}{N})+ x_0 \cdot M_{N}+T_{BCI}] + \sum_{i=1}^{N_e} \cdot [T_{RH}+T_{SH} 
\end{equation*} 
 \begin{equation}
 +  x_i \cdot M_{N}+T_{BCI}] + N_l \cdot T_{TL} \bigg]
\end{equation} 

Note that above regret is a function of $x_i$ which increases with $T$. This is unavoidable in the dynamic network as the SUs has to periodically look for any top channel vacated by the leaving SUs and need to enter into the BCI state for some time during which they incur regret. If the horizon in longer, the number of times they enter into the BCI also increases. However, if we know a priori the value of $T$, we can regulate the rate at which the SUs enter into the BCI state. For example, we set $T_{TL}=O(\sqrt{T})$, then $x=\sqrt{T}$ and we get $R_T=O(\sqrt{T})$.\hfill\IEEEQED
\begin{remark}
As the number of SUs in the network increases to the number of channels i.e. $U = N$, every SU will be selecting one of the top channels in each time slot without causing any regret and thus the expected regret of the network will lead to:
\begin{equation}
\label{t11}
R_{T} \leq N \cdot \big[ T_{RH} + x_0 \cdot T_{BCI} \big].
\end{equation}
\end{remark}

\vspace{-0.2cm}
\section{Synthetic Experiment and Analysis}\label{sim_res}
For synthetic experiments, we consider TSN and TDN algorithms, separately. For TSN algorithm, we show the comparisons with the two variants of the DLF algorithm, one with known $U$ (DLF) and other with unknown $U$ (DLF-Un) \cite{gai2}, state-of-the-art MC algorithm \cite{MC} and SH based algorithm in \cite{quek}. Since it has been shown in \cite{gai2} that the DLF algorithm outperforms the $\rho^{rand}$ and TDFS algorithms, we don't include them here to avoid repetitive results and maintain clarity of the plots. The comparison is done with respect to the parameters: 1)~Vacant spectrum utilization ($S_T$) in \%, 2) Total throughput loss, i.e., average regret, 2)~Average number of SU collisions. Later, the effect of $U$ and $N$ on $S_T$ is analyzed for the TSN algorithm. In case of TDN algorithm, we show the comparison with state-of-the-art dynamic MC (DMC) algorithm \cite{MC}.

The channel statistics of $N$ bands is given by $\mu_{\lceil \frac{N}{2}\rceil} = 0.5$ and for $n > {\frac{N}{2}}$ and $n < {\frac{N}{2}}$, the gap between the channel statistics of the $n^{th}$ and $(n+1)^{th}$ channel, i.e. $\Delta$, should be at least 0.07. For instance, for $N=8$, we set $\mu_n =  \{0.29,0.36,0.43,0.50,0.57,0.64,0.71,0.78\}$. This is referred to as Case 1. In addition, we consider second statistics for $N$=8 which is referred to as Case 2:- $\mu_n~=~\{0.10,0.20,0.30,0.40,0.50,0.60,0.70,0.80\}$.  
Each numerical result shown in the analysis is the average of the values obtained over 50 independent experiments. The simulations consider the horizon of 10000 time slots for static network and large horizon of 100000 time slots for the dynamic network. The value of $T_{CC}$ and $T_{TL}$ are set to 2000 and 200, respectively. For a fair comparison, the duration of learning phase of the MC algorithm is set to 2000 time slots.

\vspace{-0.25cm}
\subsection{Spectrum Utilization for Static Network}
First, we compare the utilization of the vacant spectrum in \%, $S_T$, of various algorithms at different instants of the horizon. In OSA, $S_T$ should be as high as possible. For illustration, we consider $N$ = 8 and $U =\{4,8\}$.

\begin{figure}[!t]
	\centering
	\subfloat[]{\includegraphics[scale=0.25]{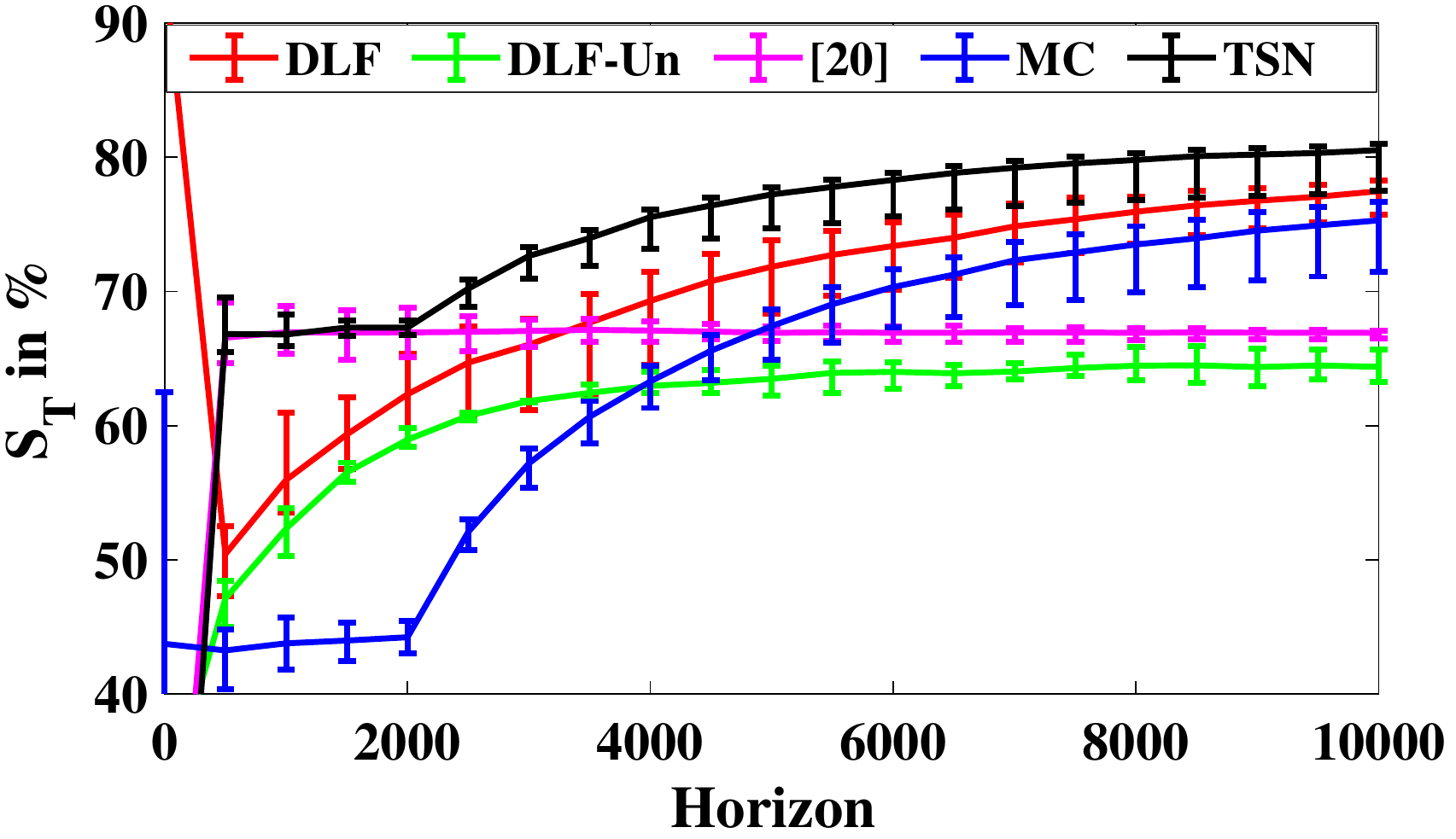}%
		\label{rd13}}
	\subfloat[]{\includegraphics[scale=0.25]{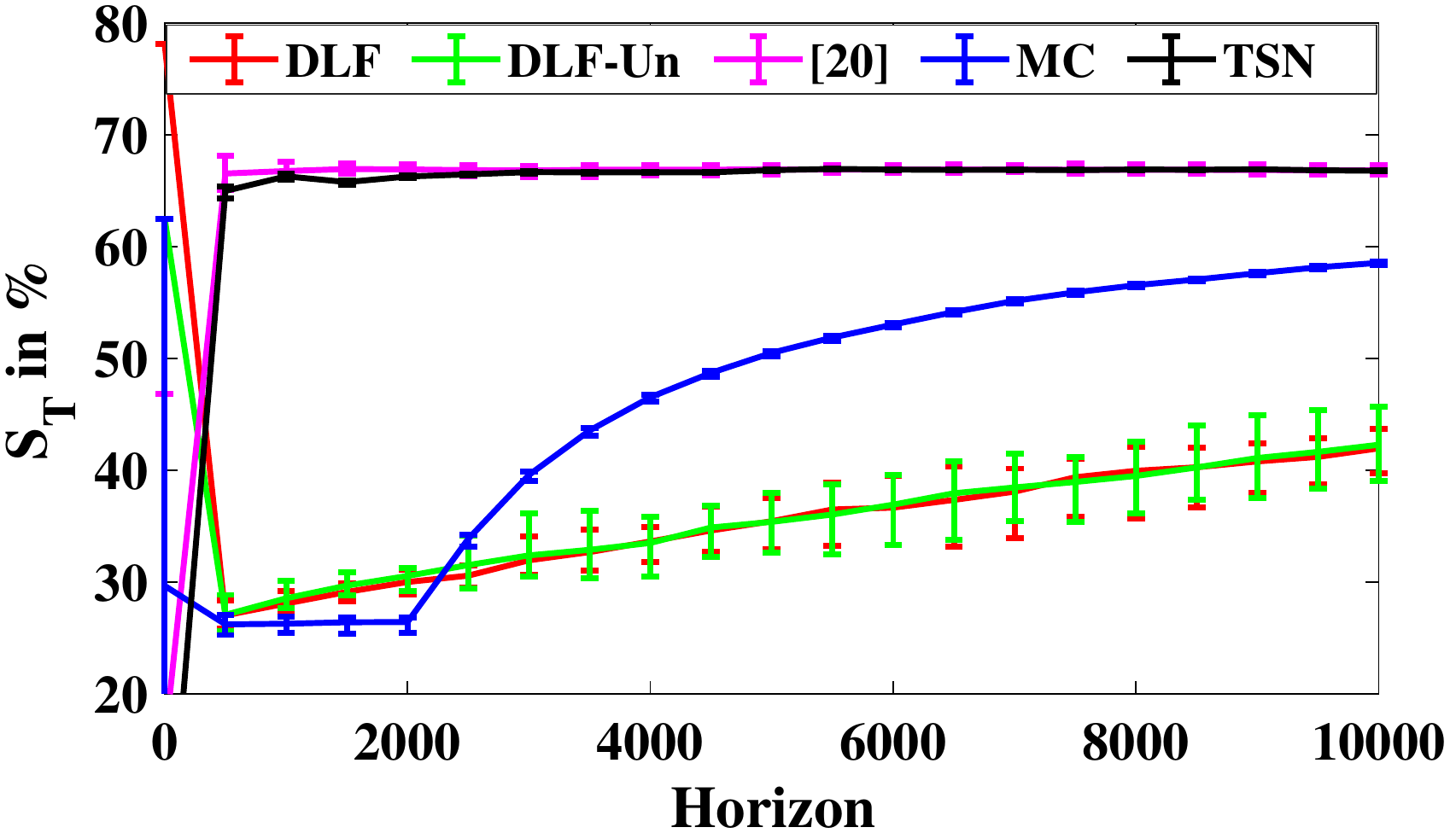}
		\label{rd14}}\\\vspace{-0.25cm}
	\subfloat[]{\includegraphics[scale=0.25]{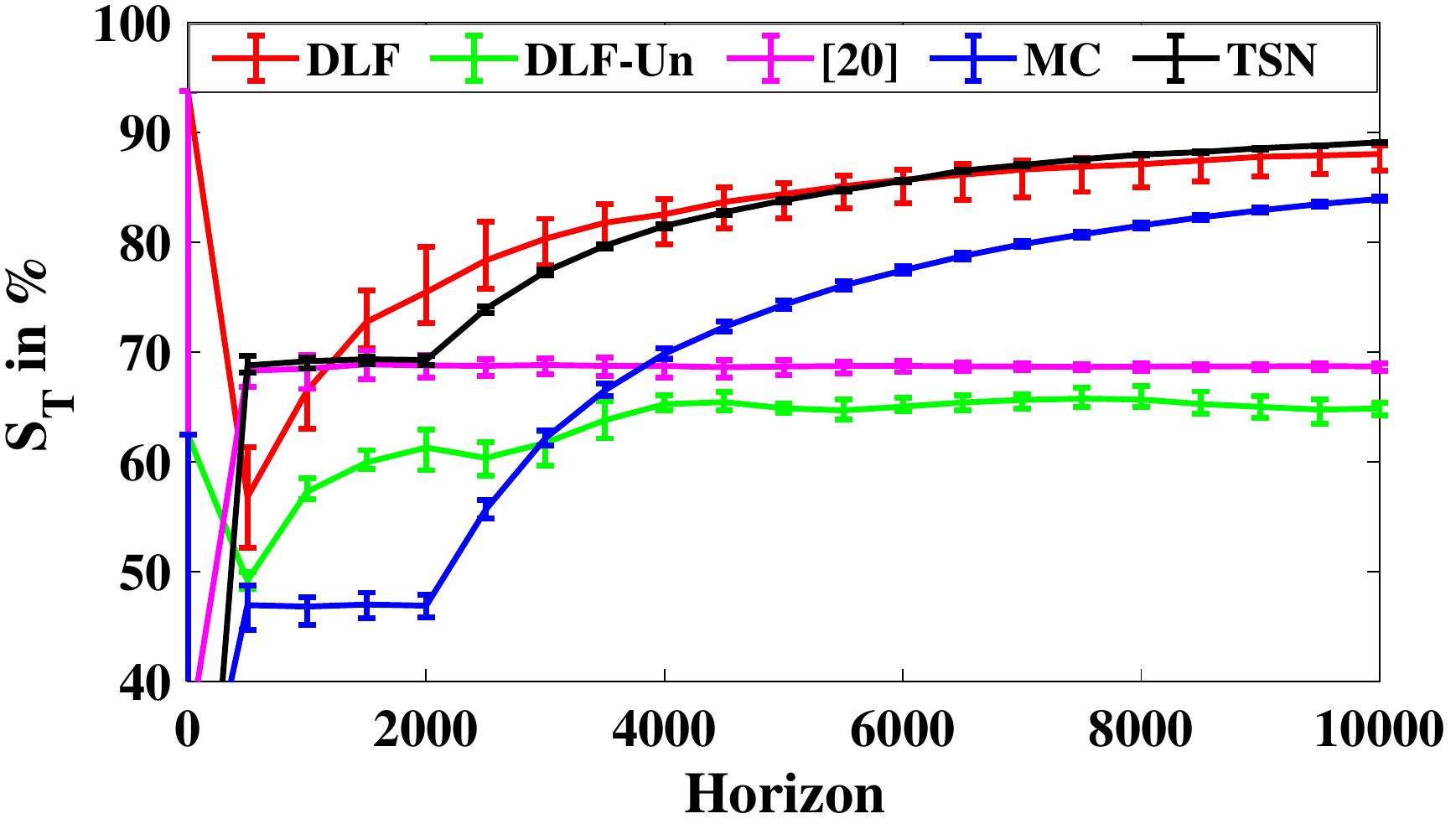}%
		\label{rd1}}
	\subfloat[]{\includegraphics[scale=0.25]{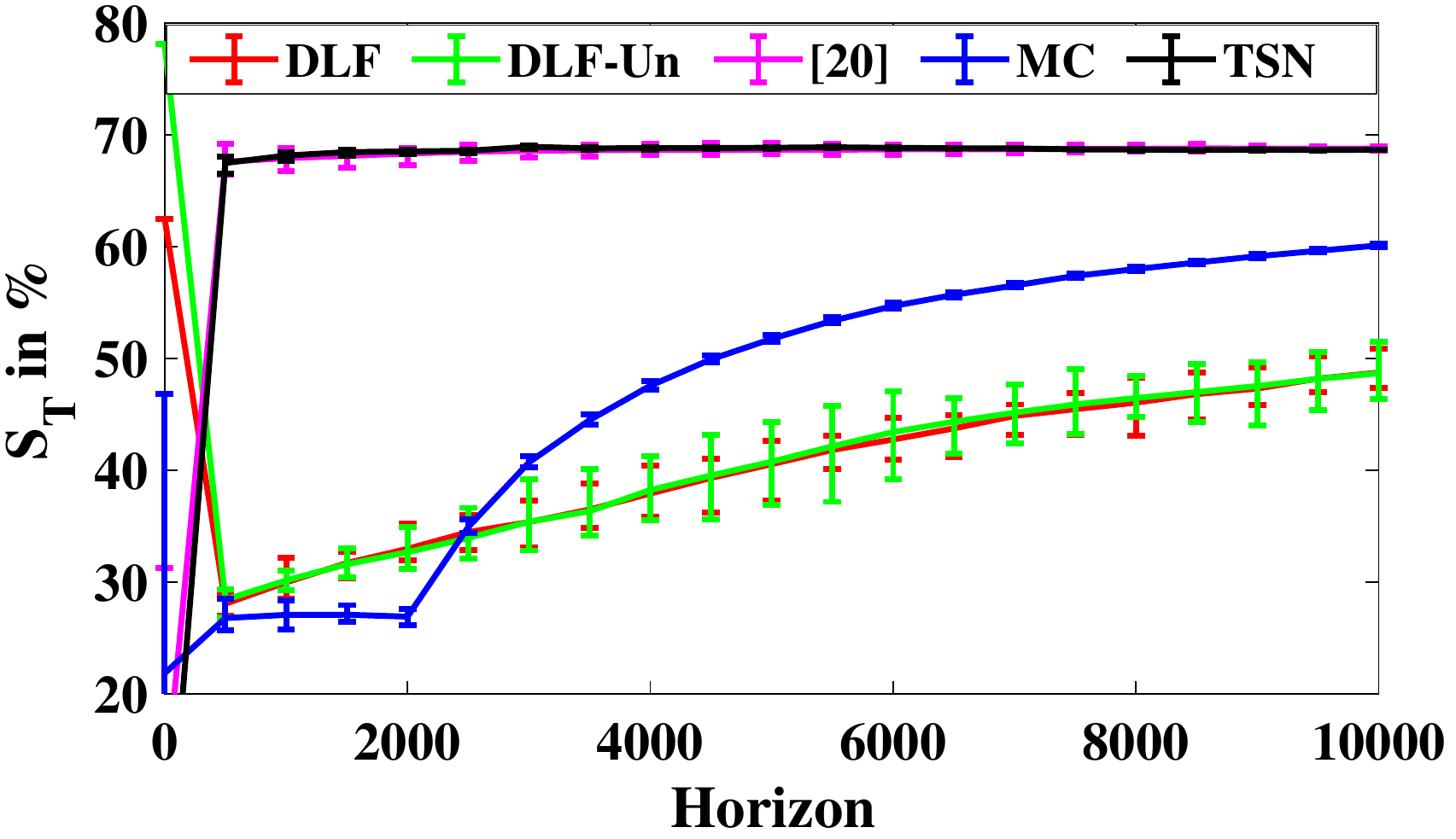}
		\label{rd2}}
	\vspace{-0.25cm}
	\caption{{\footnotesize The comparison for utilization of the vacant spectrum in \% for (a)~Case 1 with $U=4$, (b)~Case 1 with $U=8$, (c)~Case 2 with $U=4$ and (d)~Case 2 with $U=8$. Higher is better.}}
	\label{rd22}
	\vspace{-0.5cm}
\end{figure}

For statistics in Case 1, the utilization of the vacant spectrum in \% at different instants of horizon are shown in Fig.~\ref{rd13}  and Fig.~\ref{rd14} for $U$ = 4 and $U$ = 8, respectively. It can be observed that the TSN algorithm offers higher spectrum utilization than other algorithms with unknown $U$, i.e., DLF-Un, MC, and \cite{quek}. As the value of $U$ increases, the number of collisions in DLF and MC algorithms increases substantially. This does not happen in the \cite{quek} and TSN algorithms due to collision-free sequential hopping approach discussed in Theorem 1 and 2. Hence, these two algorithms perform substantially better than the DLF, DLF-Un and MC algorithms for $U$=8 as shown in Fig.~\ref{rd14}. Since the algorithm in \cite{quek} selects all channels uniformly via collision free approach, its performance is poor for smaller $U$, but improves with $U$ and is identical to the TSN algorithm when $U=N=8$. As expected, the plots for DLF and DLF-Un algorithms overlap for $U = N$.  Similar results are also observed for Case 2, and corresponding plots are shown in Fig.~\ref{rd1} and Fig.~\ref{rd2}, for $U$=4 and $U$=8, respectively.  

\begin{figure}[!b]
		\vspace{-0.25cm}
	\centering
	\subfloat[]{\includegraphics[scale=0.25]{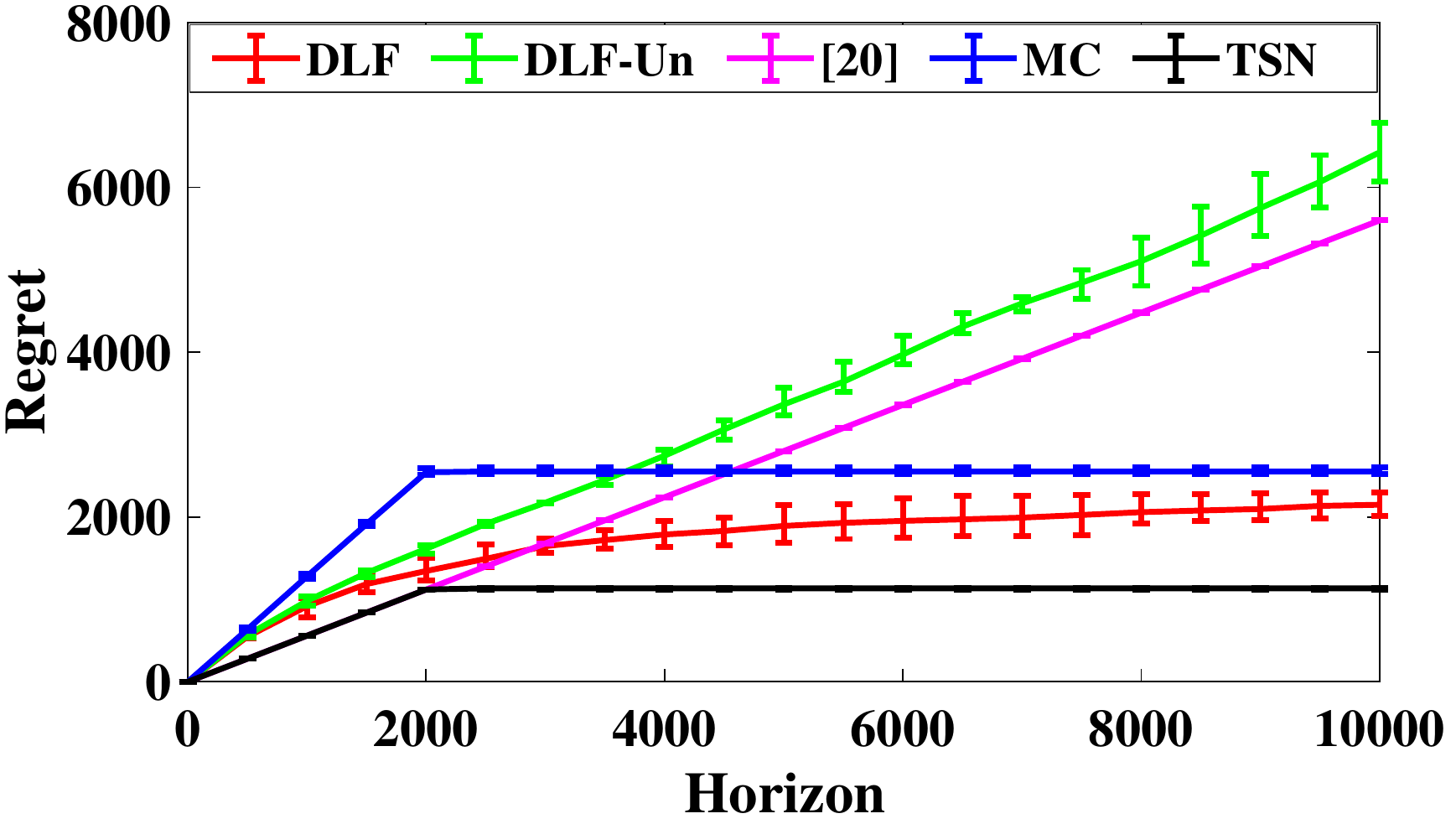}%
		\label{reg211}}
	\subfloat[]{\includegraphics[scale=0.245]{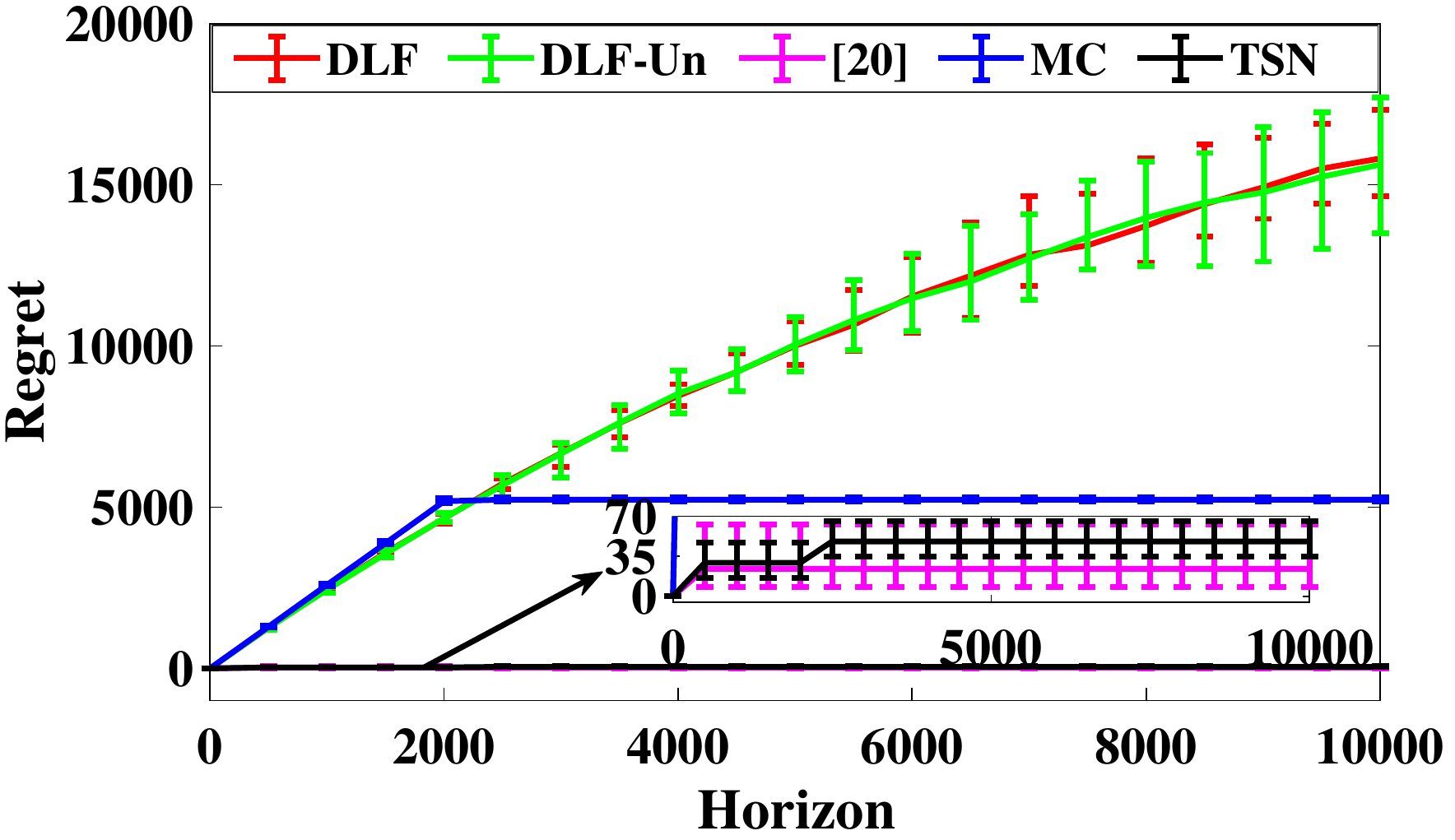}
		\label{reg222}}\\ \vspace{-0.25cm}
	\subfloat[]{\includegraphics[scale=0.255]{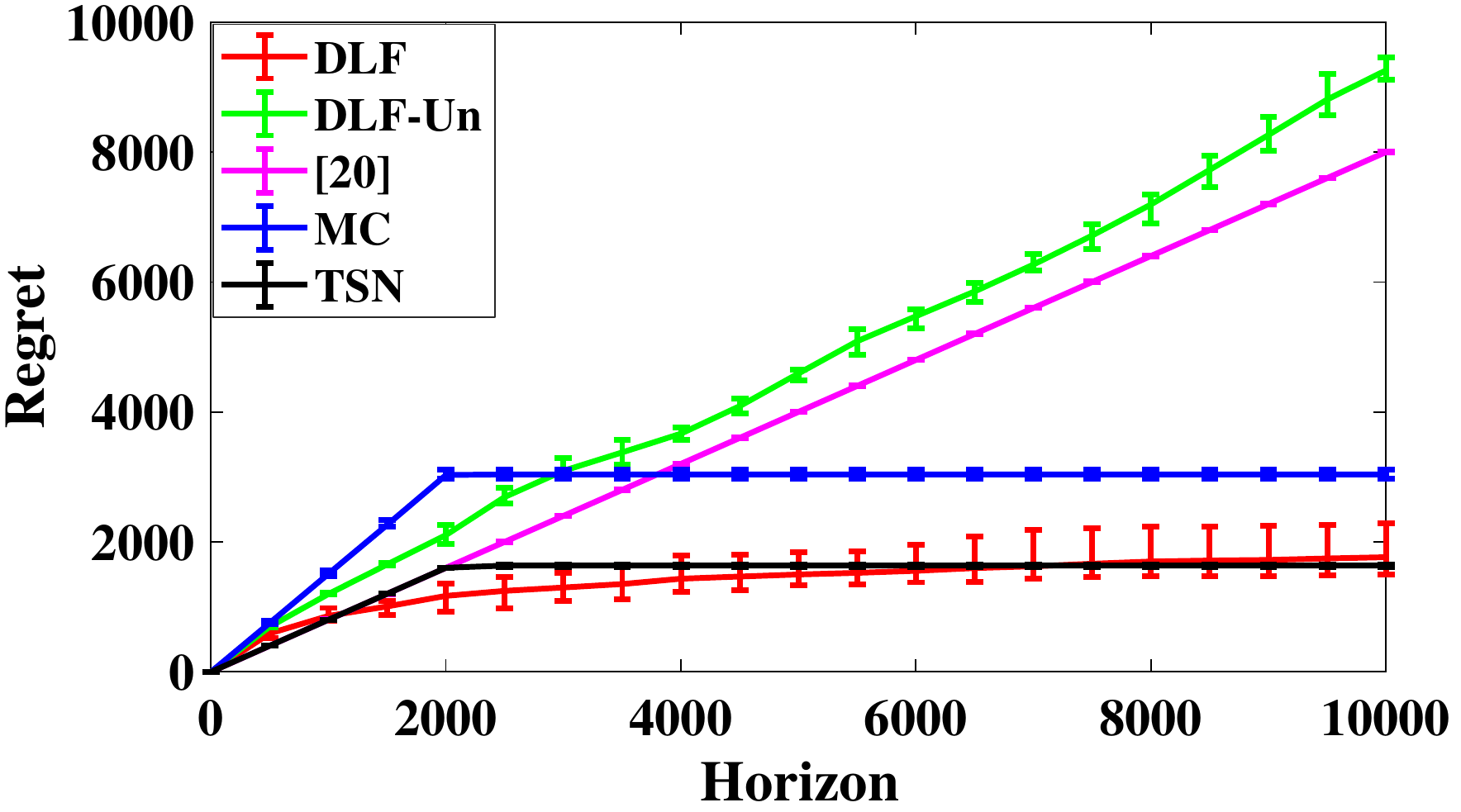}%
		\label{reg11}}
	\subfloat[]{\includegraphics[scale=0.25]{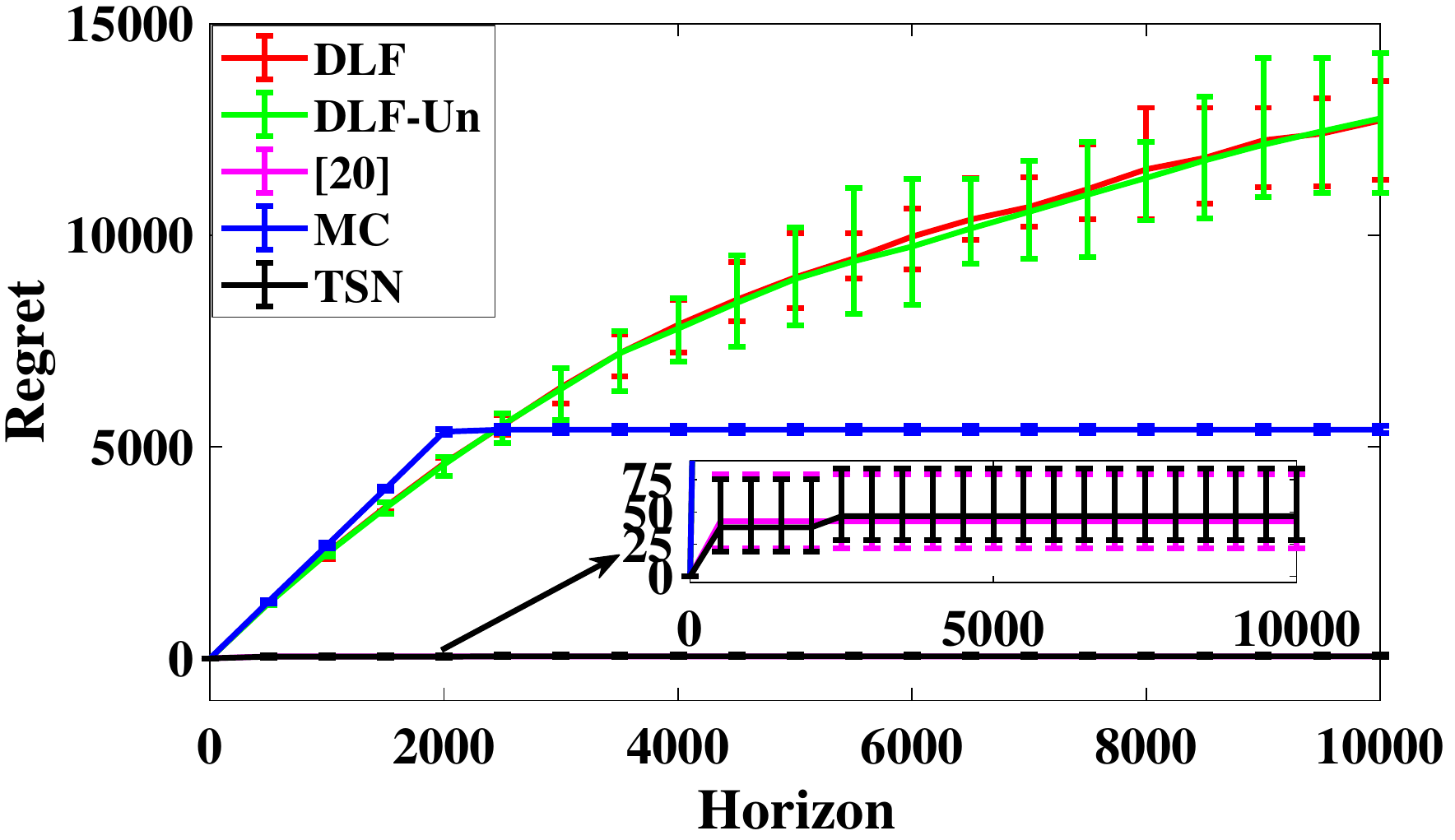}
		\label{reg12}}
	\vspace{-0.25cm}
	\caption{{\footnotesize The comparison for average regret of various algorithm for (a)~Case 1 with $U=4$, (b)~Case 1 with $U=8$, (c)~Case 2 with $U=4$ and (d)~Case 2 with $U=8$. Lower is better.}}
	\label{reg2}
	\vspace{-0.25cm}
\end{figure}

Next, we compare the average regret which is calculated using the Eq.~\ref{eq:regret}. Consider the plots in Fig. \ref{reg211} and Fig. \ref{reg222} corresponding to Case 1 with $U=4$ and $U=8$, respectively. Similar plots for Case 2 are shown in Fig. \ref{reg11} and Fig. \ref{reg12}. The constant regret (i.e. the plot with zero slope) guarantees that the SUs have settled in the top channels and hence, zero regret after that. It can be observed that the TSN algorithm significantly outperforms other algorithms and offers constant regret similar to state-of-the-art MC algorithm.

\begin{figure}[!b]
	\vspace{-0.4cm}
	\centering
	\subfloat[]{\includegraphics[scale=0.265]{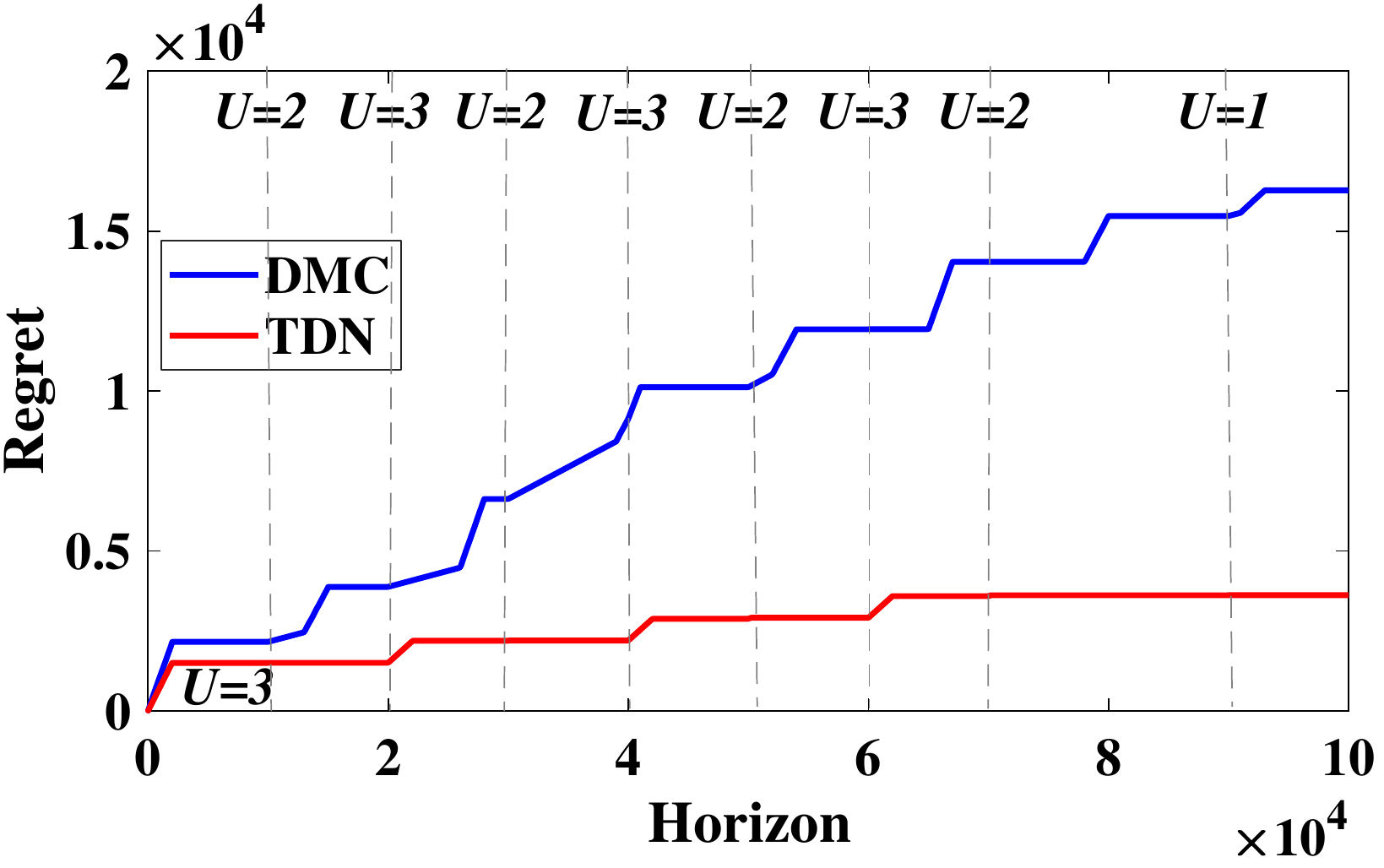}	}	
	\subfloat[]{\includegraphics[scale=0.25]{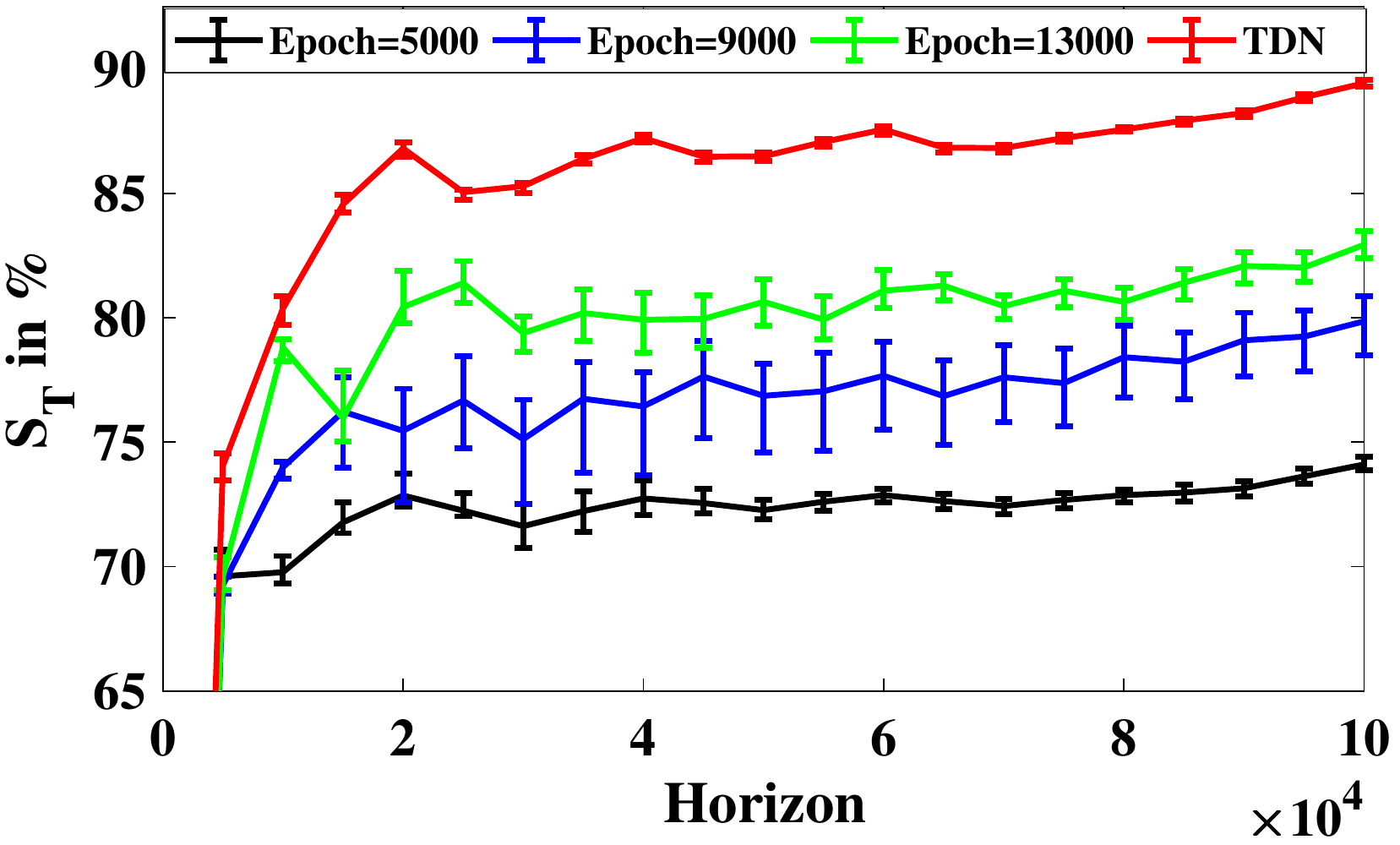}} \\\vspace{-0.45cm}
	\subfloat[]{\includegraphics[scale=0.265]{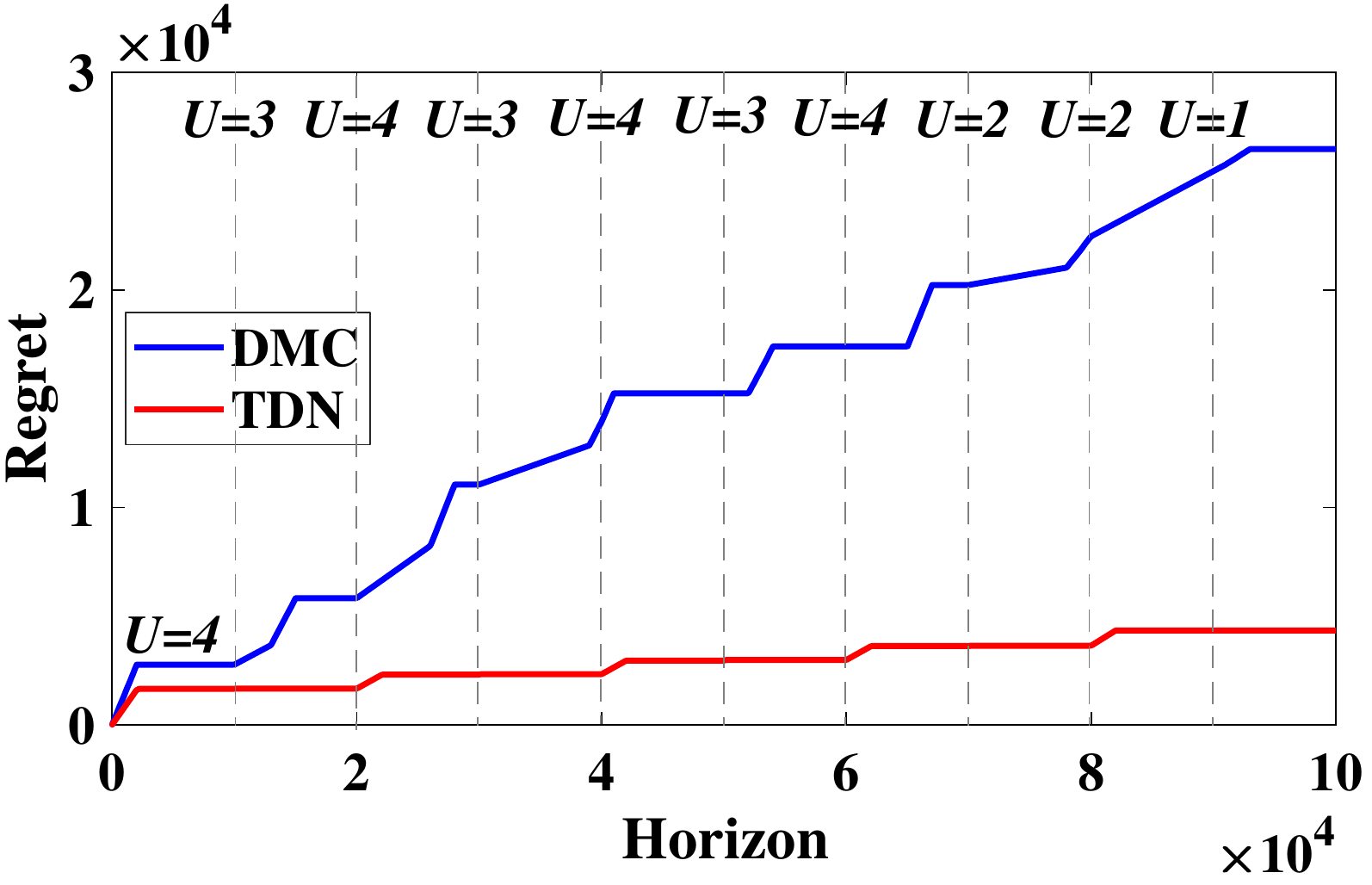}}	
	\subfloat[]{\includegraphics[scale=0.25]{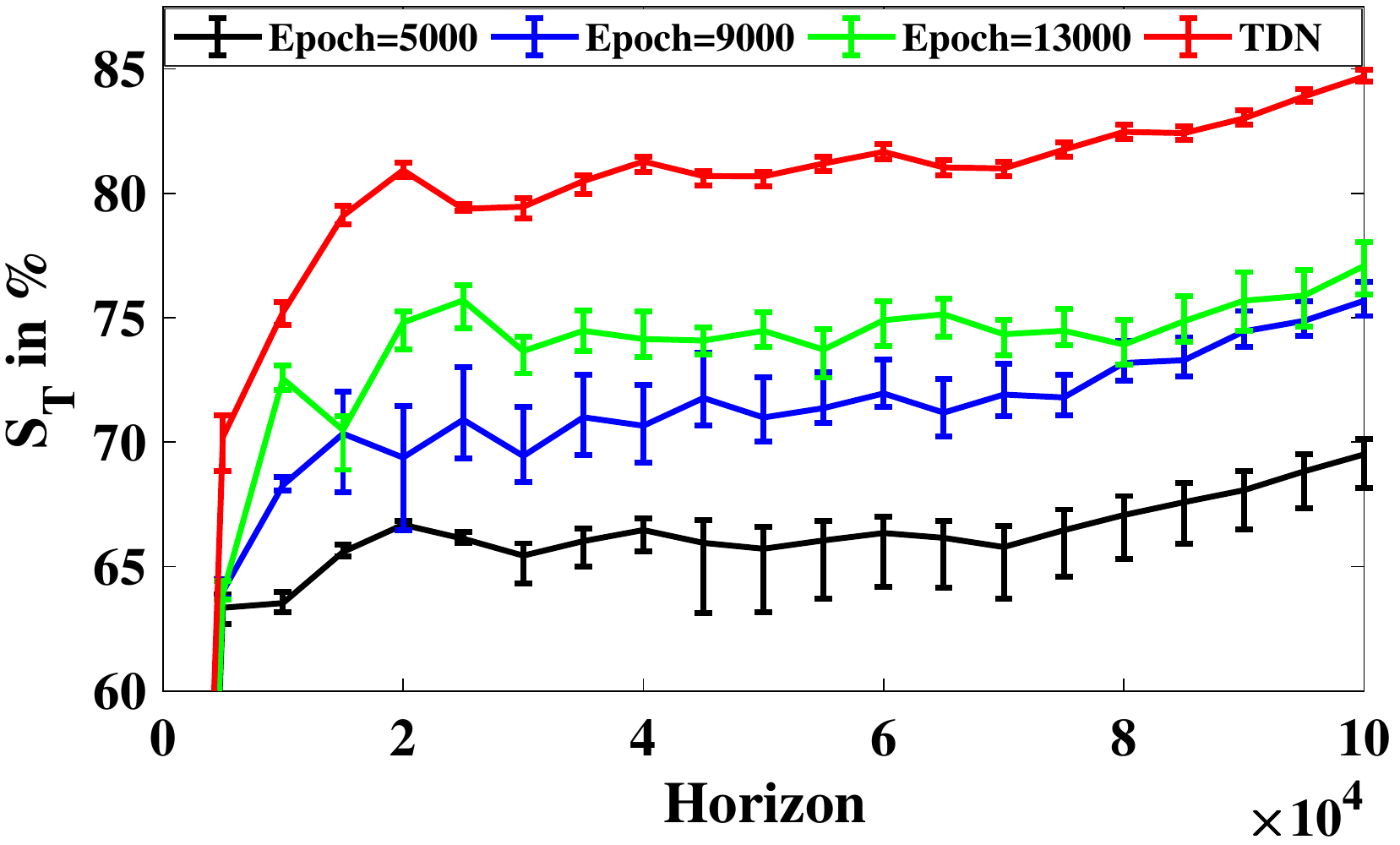}}\\ \vspace{-0.45cm}
	\subfloat[]{\includegraphics[scale=0.265]{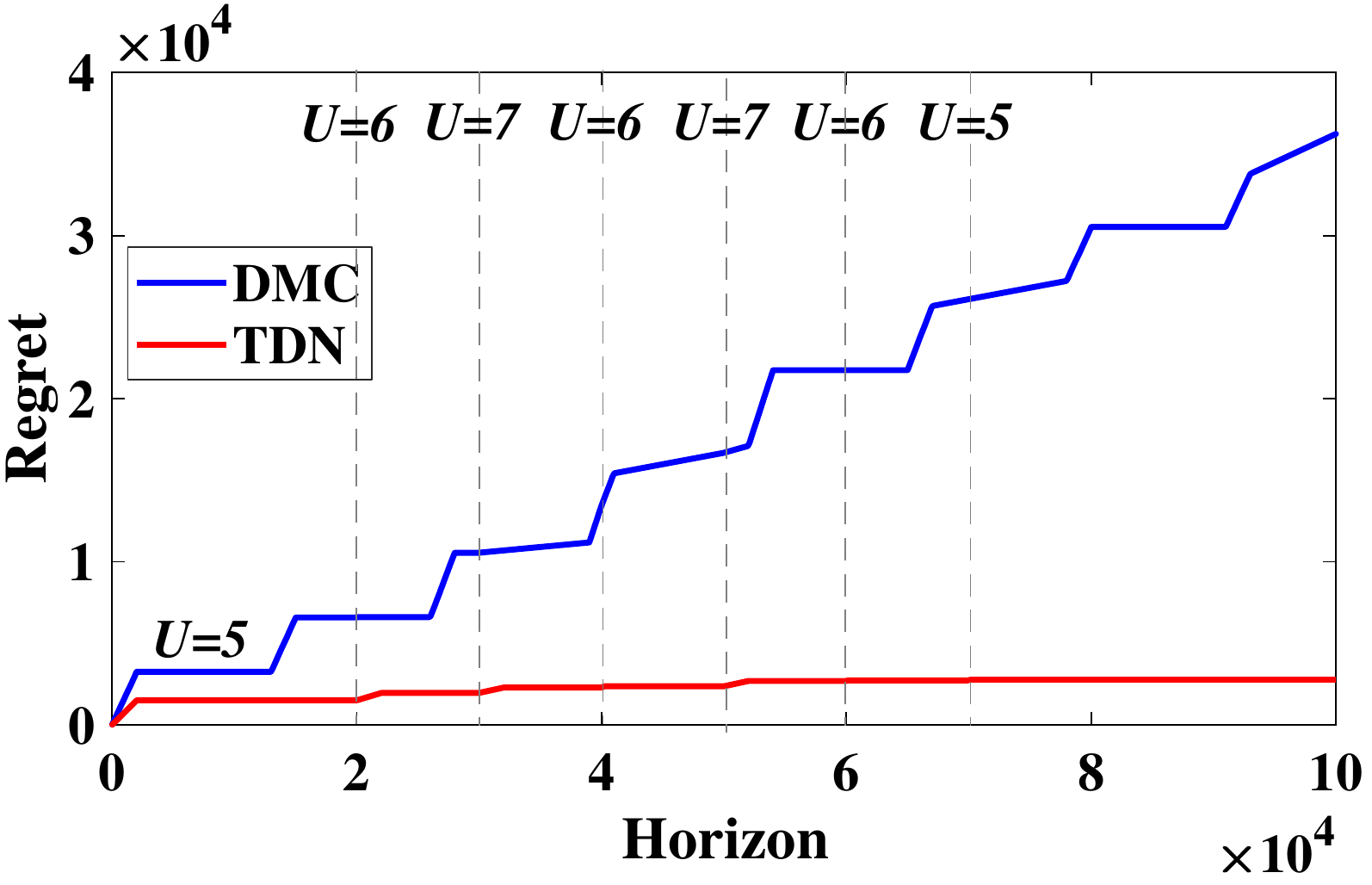}}		
	\subfloat[]{\includegraphics[scale=0.25]{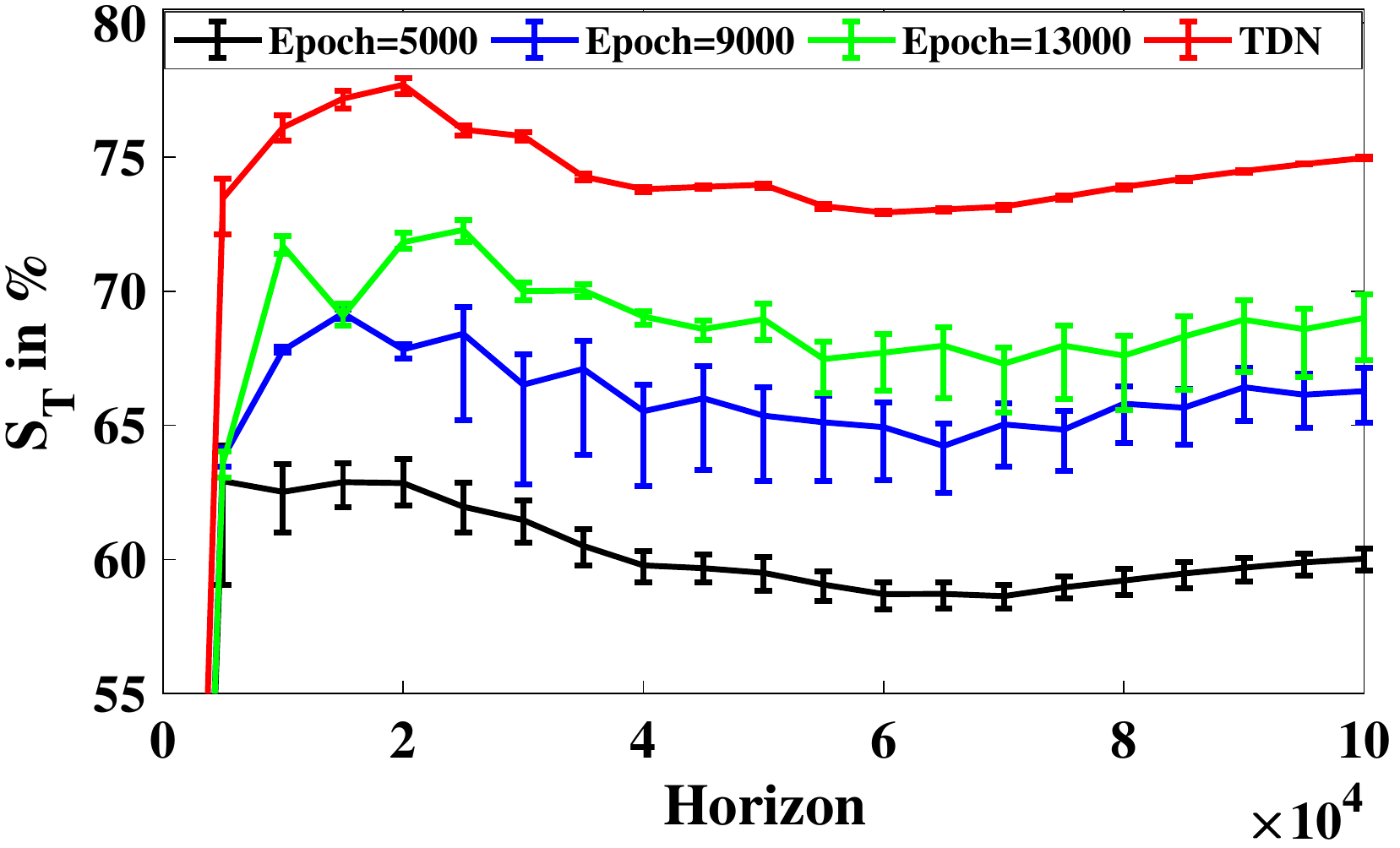}}
\vspace{-0.25cm}
	\caption{{\footnotesize Average regret of the TDN and DMC algorithms for (a) Scenario 1, (c) Scenario 2 and (e) Scenario 3. Total vacant spectrum utilization of the TDN and DMC algorithms for (b) Scenario 1, (d) Scenario 2 and (f) Scenario 3. Three different epoch length are considered for DMC while $T_{TL}$ for the TDN algorithm is 200.}}
	\label{dyn1}
\end{figure}
\vspace{-0.25cm}
\subsection{Spectrum Utilization for Dynamic Network}
In this section, we compare the performance of the TDN algorithm with the state-of-the-art DMC algorithm \cite{MC}. Here, we consider the large horizon of 100000-time slots and three different scenarios depicting the various combination of the time interval at which the SUs enter or leave the network. Each numerical result shown in the analysis is the average of the values obtained over 50 independent experiments. For illustration, we consider $N$ = 8 with channel statistics given as:- $\mu_n =  \{0.29,0.36,0.43,0.50,0.57,0.64,0.71,0.78\}$.

Consider the average regret and spectrum utilization plots for Scenario 1 shown in Fig.~\ref{dyn1} (a) and (b), respectively. We have indicated the number of active SUs at different instants of the horizon in Fig.~\ref{dyn1} (a). For example, there are 3 SUs in the beginning. Then, one SU leaves at $t=10000$, one SU enters at $t=20000$ and so on. Similarly, two more scenarios are considered in the rest of sub-figures of Fig.~\ref{dyn1}. The value of $T_{TL}$ is set as 200 and the epoch length of the DMC \cite{MC} algorithm is carefully chosen as 13000 as it offers the best performance in each of the three scenarios.

%

As discussed in Section~\ref{DN}, the regret of the DMC algorithm is significantly higher than that of the TDN algorithm. This is because the DMC algorithm incurs regret which is linear in time during the learning phase and this phase repeats in every epoch. On the other hand, the TDN algorithm follows epoch-less trekking and collision-free sequential hopping approach. This also means that the vacant spectrum utilization of the TDN algorithm is higher than the DMC algorithm. Similar behavior can also be observed in Fig.~\ref{dyn1}.

\par

\vspace{-0.25cm}

\subsection{Number of SU Collisions}
\begin{figure}[!h]
	\centering
	\subfloat[]{\includegraphics[scale=0.25]{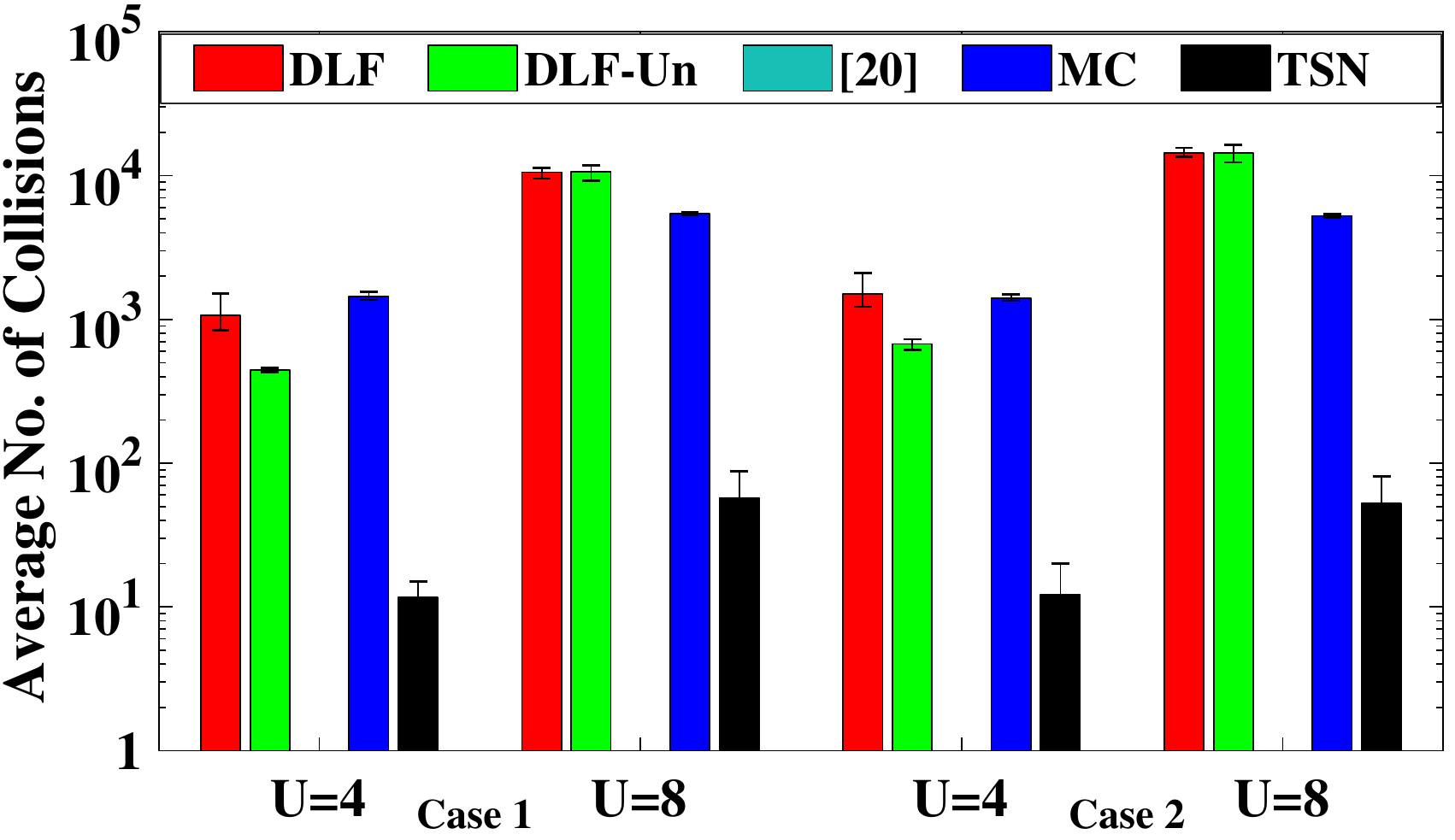}%
		\label{rd3}} \hspace{0.01cm}
	\subfloat[]{\includegraphics[scale=0.26]{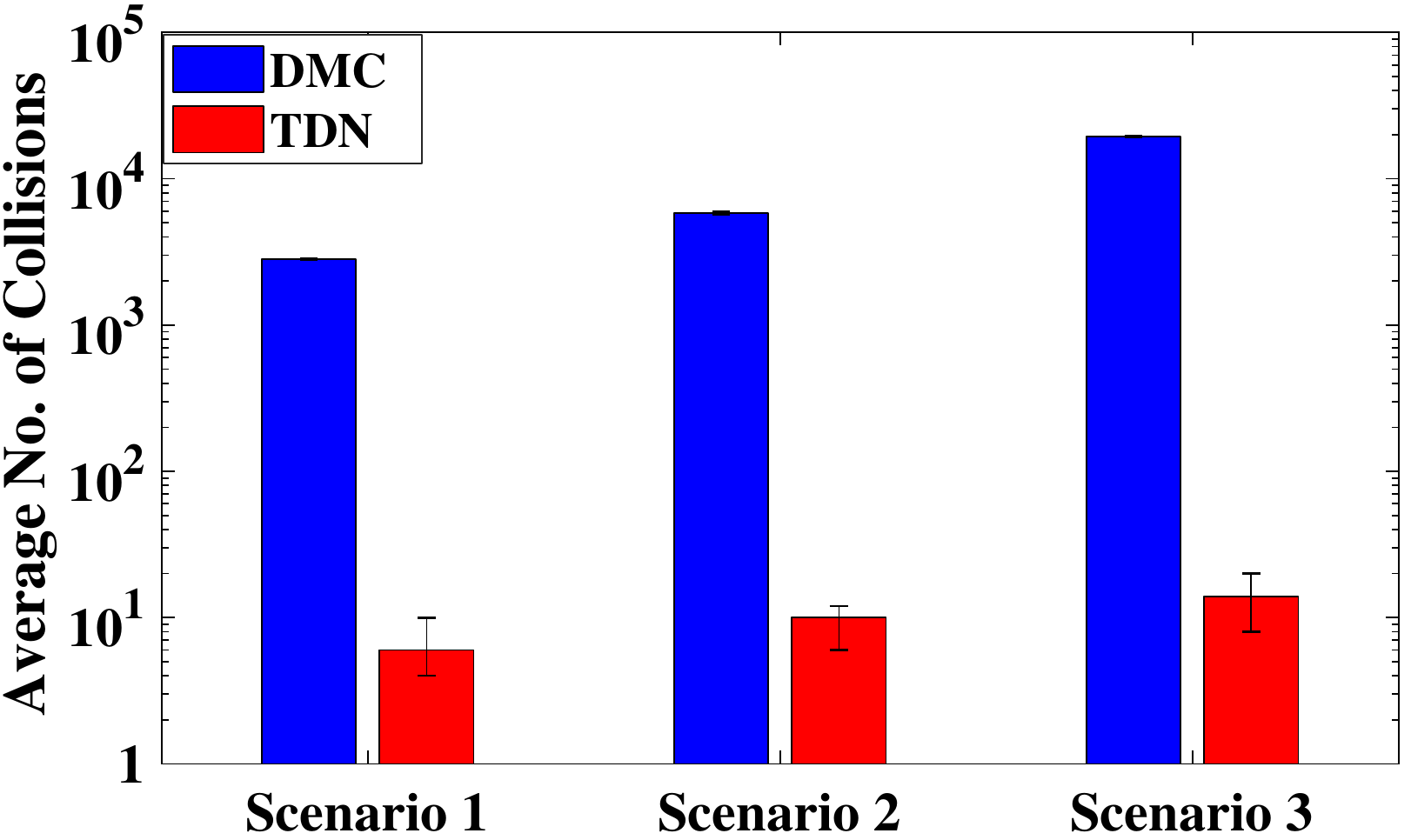}
		\label{rd5}}
	\vspace{-0.25cm}
	\caption{{\footnotesize (a) Average no. of collisions for Case 1 and Case 2, with $U=4$ and $U=8$ for static network, and (b) Average no. of collisions for three different scenarios considered in dynamic network. Note that $y-axis$ is represented on a logarithmic scale.}}
	\label{colls}
\end{figure}

We compare the average number of collisions, $C_T$, faced by all SUs  at the end of the horizon for static and dynamic networks. The corresponding plots are shown in Fig.~\ref{colls} with the data on $y-axis$ represented on the logarithmic scale for better visualization. It can be observed that the number of SU collisions are negligible in the \cite{quek} and TSN algorithms. The number of SU collisions are close to zero in \cite{quek} compared to at most 50 collisions in the TSN algorithm. This corresponds to very small collision probability of 0.001 in the TSN algorithm. However, the vacant spectrum utilization and regret of \cite{quek} is significantly poor than the TSN algorithm as discussed in previous section making the TDN algorithm preferred choice over the algorithm in \cite{quek}. For dynamic networks comparison in Fig.~\ref{rd5}, the number of collisions in the TDN algorithm is negligible due to the collision-free hopping and epoch-less approach compared to random hopping and epoch based DMC algorithm. 

%

To summarize, we have compared the performance of the TSN and TDN algorithms with the existing state-of-the-art algorithms. It can be observed that the proposed algorithms outperform existing algorithms in terms of average regret, vacant spectrum utilization and number of collisions. In addition, the proposed algorithms are simple to design and implement and does not need any complex algorithms like \cite{prand,gai2}. Fewer number of collisions lead to fewer packet reprocessing and fewer transmissions leading to saving in the power consumption. This makes the TSN and TDN algorithms preferred choice for the battery operated SU terminals.

\vspace{-0.2cm}
\section{Experimental Results and Analysis}\label{exp_res}
The simple USRP based testbed has been developed to validate the functionality of the proposed algorithms in real radio environment and compare it with existing state-of-the-art algorithms. The testbed consists of primary user traffic generator designed using OFDM based transmitter realized in LabView and USRP-2922 from National Instruments for over-the-air transmission. It transmits the signal in multiple channels based on their statistics. The first channel is dedicated for synchronization, and hence, it is not used by the PUs/SUs for their transmission. The synchronization has been achieved by switching the corresponding channel from occupied to vacant states or vice-versa in each time slot. Each slot duration ($\Delta t$) is 0.1 second which can be changed as per the requirement. For experiments, the transmission parameters such as the number of OFDM sub-carriers, number of channels, center frequency, and bandwidth are 1024, 8, 935 MHz and 2 MHz, respectively. At the receiver side, SUs are implemented using MATLAB and USRP N200 from Ettus Research. At each SU, the channel selected by the underlining algorithm is passed through non-ideal energy detector to check whether it is vacant or occupied. When the channel is vacant, and it is not selected by other SUs, it is assumed that the SU transmits over the channel and transmission is successful. 


 We consider $N$ = 8 with $\mu$= [0.10,0.20,0.30,0.40,0.50,0.60,0.70,0.80] and $U$ = 4 and $U$ = 8. Rest of the parameters are same as simulation results presented in previous sub-section. Each numerical result shown in the analysis is the average of the values obtained over 10 independent experiments in real radio environment and simulations consider a time horizon of 7500 slots.

\begin{figure}[!b]
	\vspace{-0.4cm}
	\centering
	\subfloat[]{\includegraphics[scale=0.25]{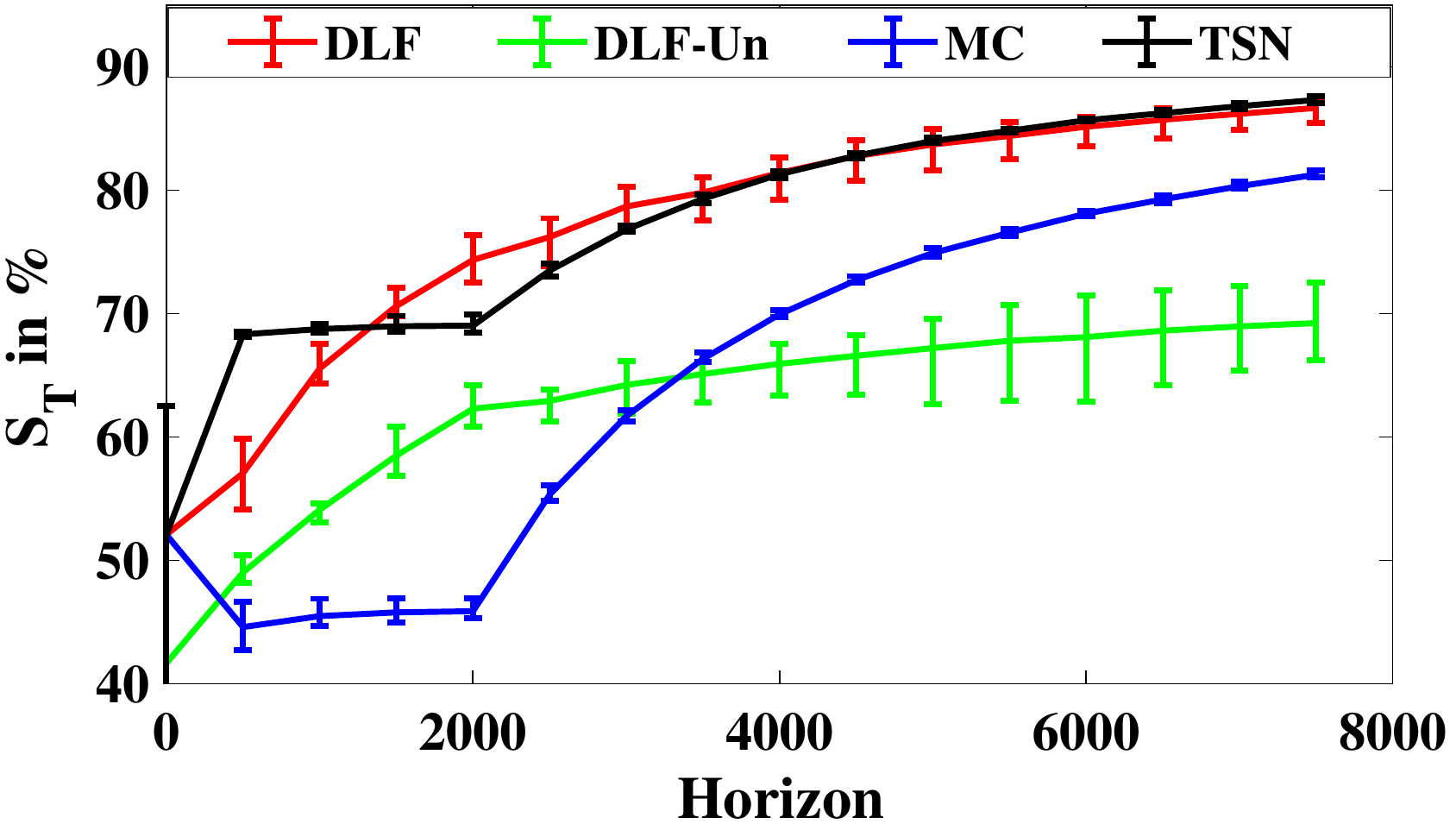}%
		\label{reg21}}
	\subfloat[]{\includegraphics[scale=0.25]{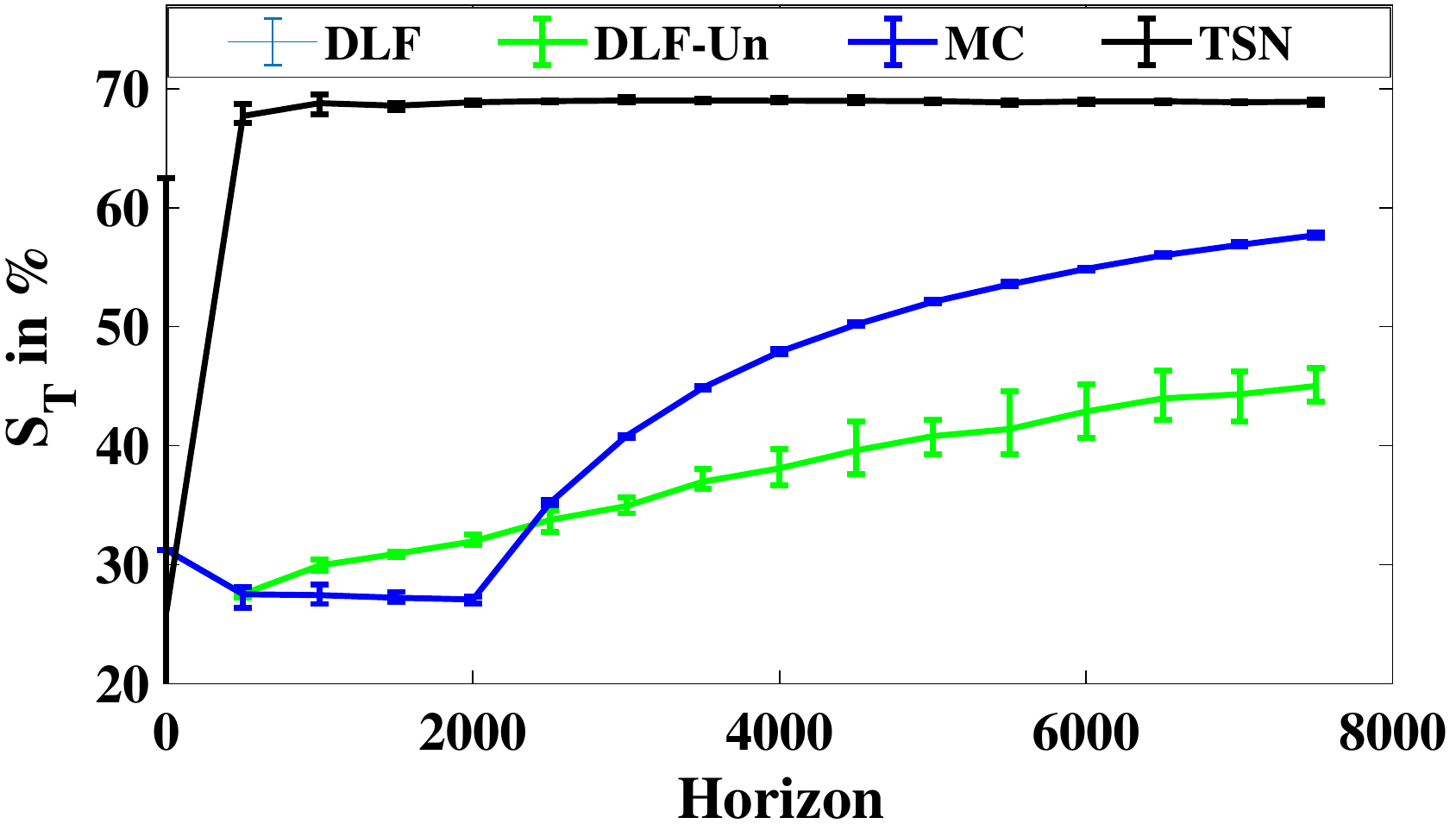}
		\label{reg22}}\\ \vspace{-0.25cm}
	\subfloat[]{\includegraphics[scale=0.26]{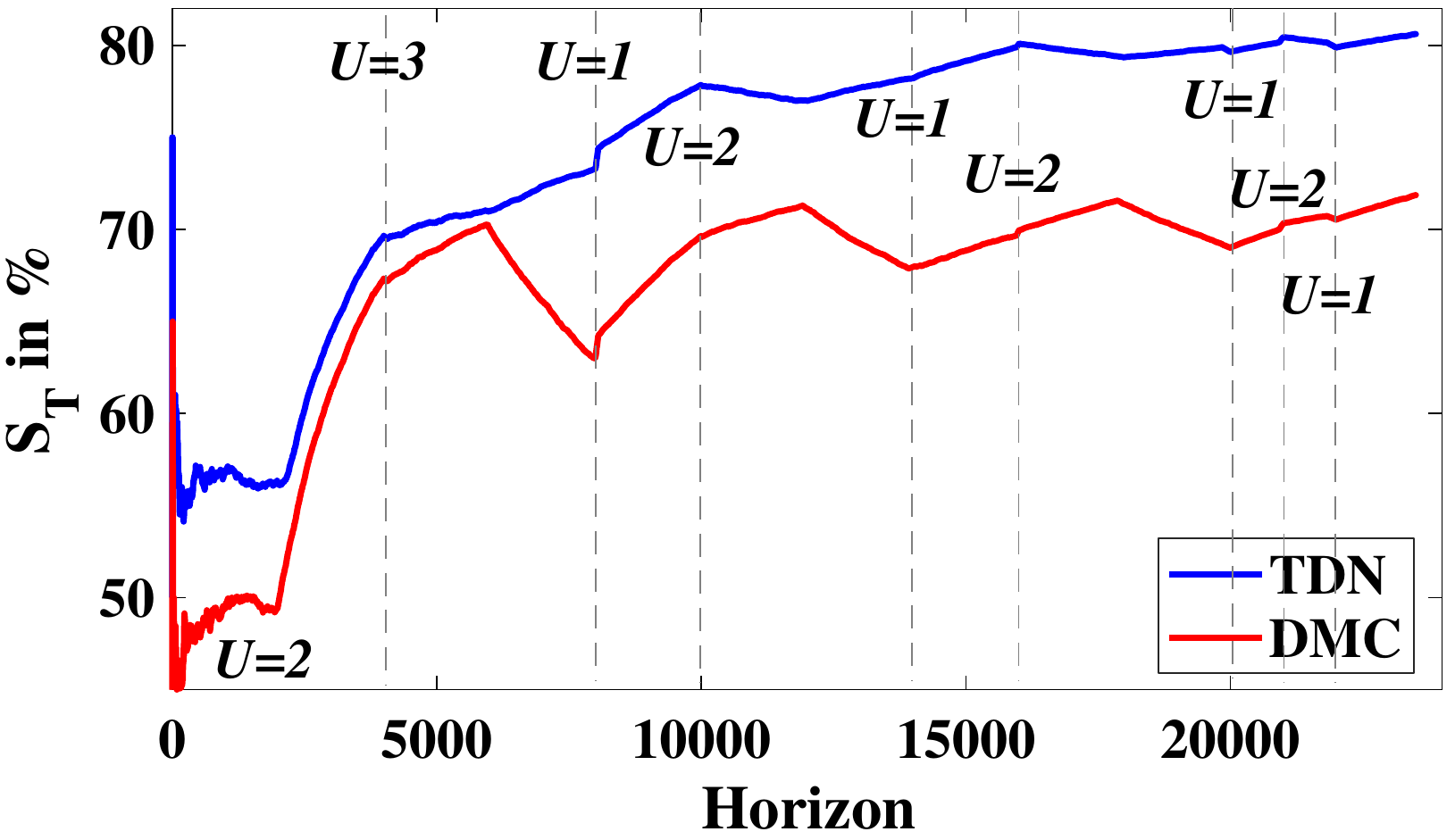}%
		\label{usrp_dyn1}}
	\subfloat[]{\includegraphics[scale=0.26]{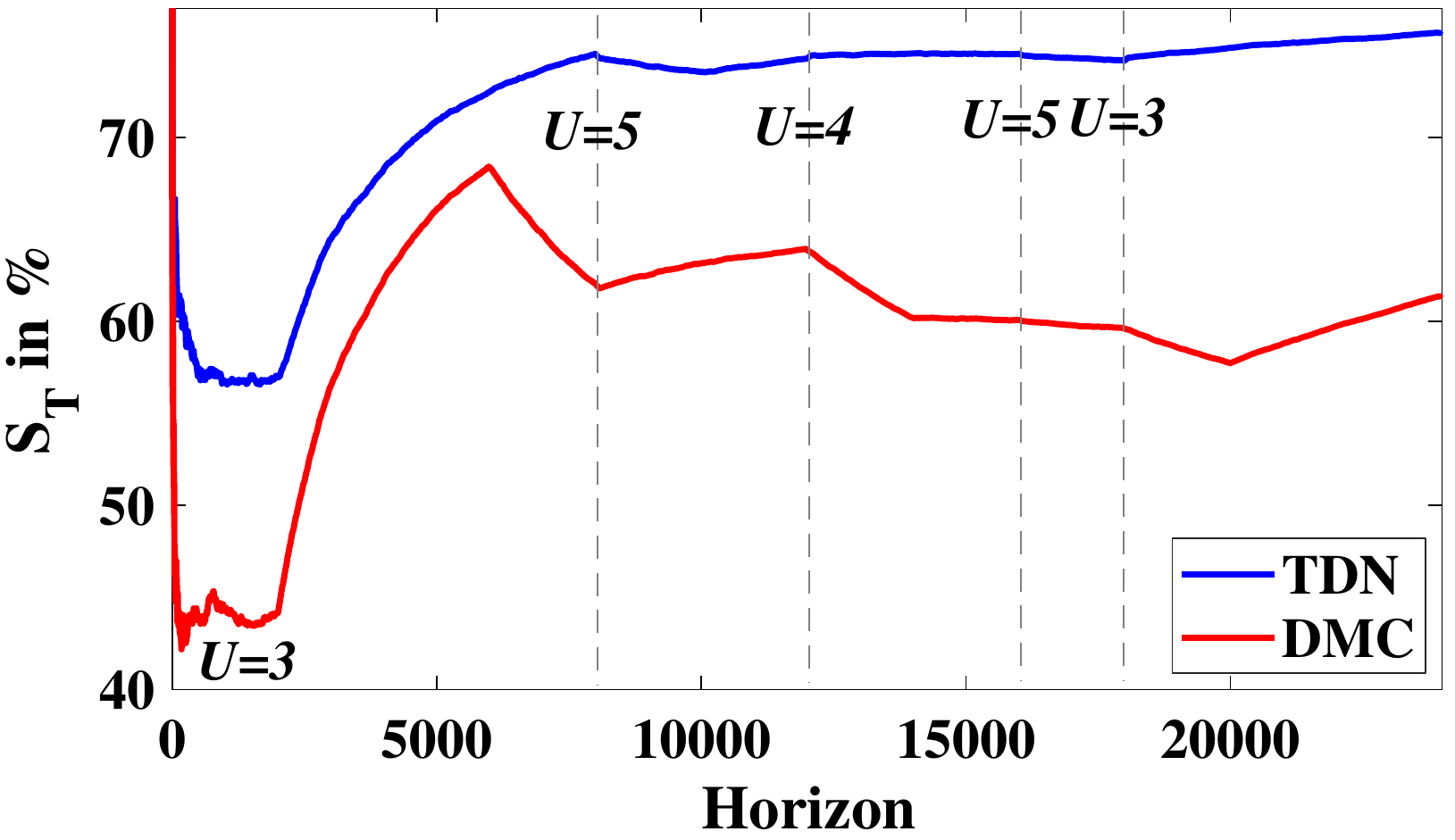}
		\label{usrp_dyn2}}
\vspace{-0.25cm}
	\caption{{\footnotesize Comparison for output reward for static network TSN in \% for a) $U$ = 4 and  b) $U$ = 8. Comparison for output reward of TDN in \% for c) Scenario 1 and  d) Scenario 2.}}
	\label{reg2u}
\vspace{-0.45cm}
\end{figure}

We now compare the proposed TSN algorithm with the DLF algorithm \cite{gai2}, DLF-Un algorithm and MC algorithm \cite{MC}. In Fig.~\ref{reg2u}, we compare the total average spectrum utilization in \% of these algorithms for $U=4$ and $U=8$. It can be observed that the performance of the TSN algorithm approaches to that of DLF algorithm \cite{gai2} which has prior knowledge of $U$. Also, TSN algorithm significantly outperforms other algorithms which do not have prior knowledge of $U$. The results validate the simulation results presented in previous Section though actual values may differ due to sensing errors in real radio environment.


%

\begin{figure}[!b]
	\vspace{-0.5cm}
	\centering
	\subfloat[]{\includegraphics[scale=0.25]{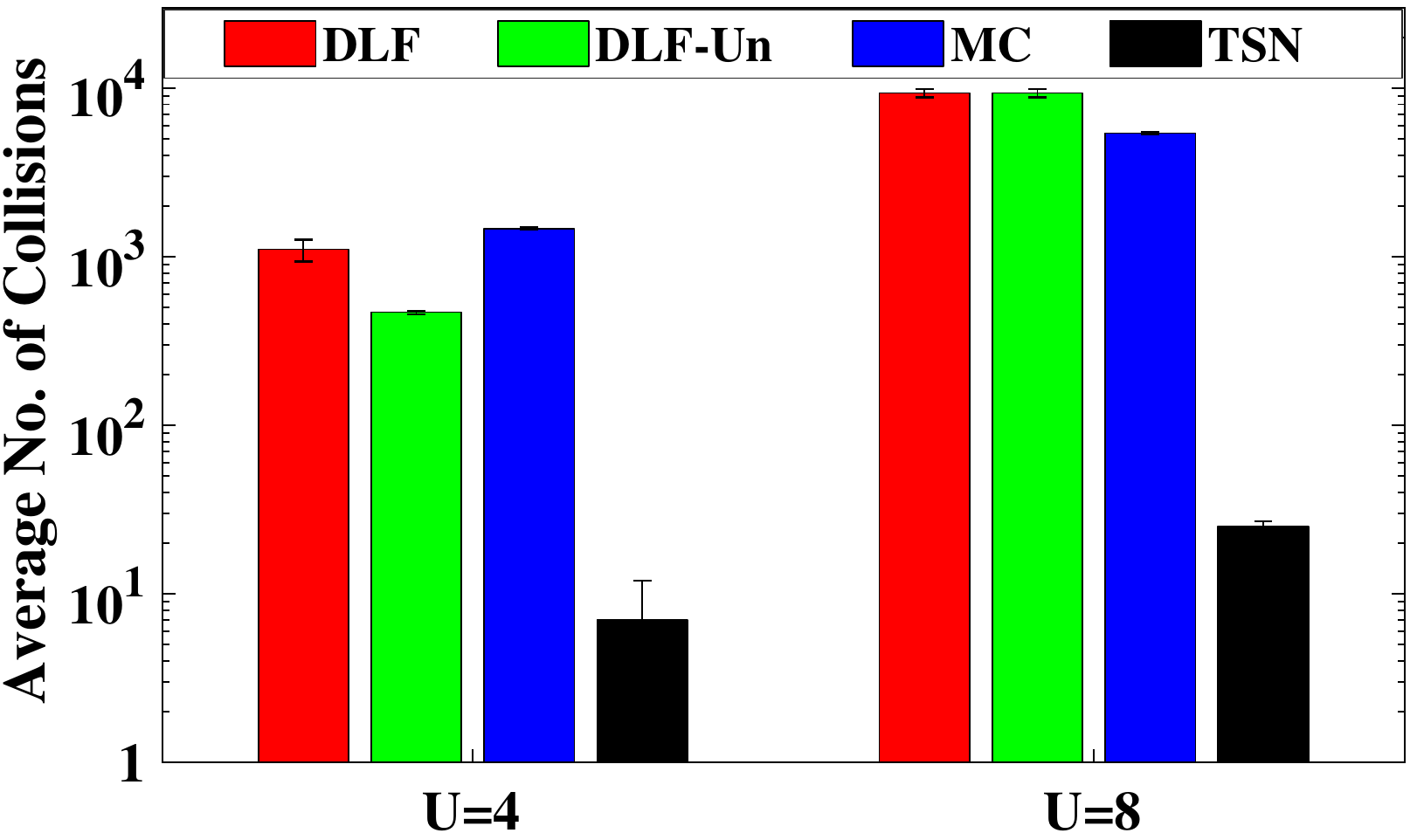}%
		\label{rd611}}
	\subfloat[]{\includegraphics[scale=0.25]{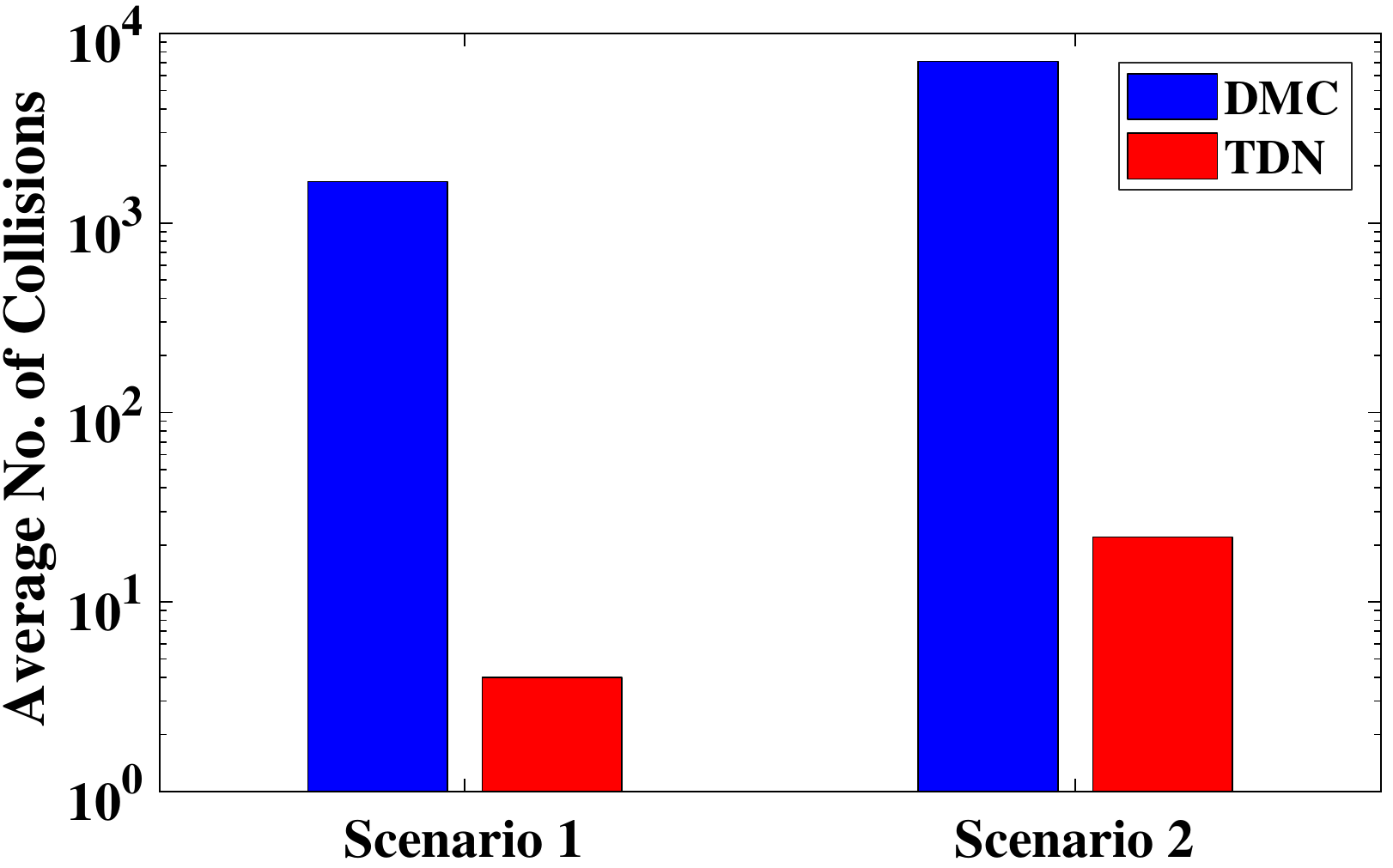}
		\label{usrp_dyn3}}
	\vspace{-0.2cm}
	\caption{{\footnotesize (a) Average no. of collisions for $N$ = 8 with $U$ = 4 and  $U$ = 8, (b) Average no. of collisions two scenarios.}}
	\label{usrp_dyn}
	\vspace{-0.15cm}
\end{figure}

Now we compare the performance of the TDN algorithm with the state-of-the-art DMC algorithm \cite{MC} for dynamic network. Here, we consider the horizon of size 24000 time slots and two different scenarios depicting the various combination of time interval at which the SUs enter or leave the network.  The value of $T_{TL}$ is fixed as 200 and for DMC algorithm, we choose best possible epoch length of 6000 time slots. The plots in Fig.~\ref{usrp_dyn1} and Fig.~\ref{usrp_dyn2} confirm the superiority of the proposed TDN algorithm over the DMC algorithm in real radio environment. As shown in Fig.~\ref{usrp_dyn}, the number of collisions are fewer in the proposed algorithms. Higher vacant spectrum utilization and fewer number of SU collisions in synthetic as
well as experimental results along with the theoretical analysis validate the superiority of the proposed TSN and TDN algorithms over
existing state-of-the-art algorithms.

\vspace{-0.2cm}
\section{Conclusions and Future Works} \label{conclusion}
In this paper, we proposed novel algorithms for opportunistic spectrum access (OSA) in infrastructure less cognitive radio network. The proposed algorithms are based on trekking approach where each SU continuously trek towards the better channels without knowing the number of active users in the network. The simulated and experimental results for the static as well as dynamic networks show that the proposed algorithms outperform existing algorithms in terms of vacant spectrum utilization, regret and the number of collisions.  In future, we would like to extend the proposed trekking approach for the scenario where some SUs are not faithful and may deviate from a given algorithm. 
\vspace{-0.32cm}

\ifCLASSOPTIONcaptionsoff
  \newpage
\fi

\end{document}